\documentclass[11pt,preprint2]{aastex701}
\graphicspath{{./}{}}
\usepackage{float}
\usepackage[caption=false]{subfig}
\usepackage{amsmath}

\begin{document}
\title{Overmassive supermassive black holes in SDSS close galaxy pairs}
\author[0000-0001-7829-4764]{Maggie C. Huber}
\affiliation{University of Colorado Boulder, Boulder, CO 80309, USA}
\email[show]{margaret.huber@colorado.edu}
\author[0000-0001-8627-4907]{Julia M. Comerford}
\affiliation{University of Colorado Boulder, Boulder, CO 80309, USA}
\email{julie.comerford@Colorado.EDU}
\author[0000-0003-1407-6607]{Joseph Simon}
\affiliation{University of Colorado Boulder, Boulder, CO 80309, USA}
\altaffiliation{NSF Astronomy \& Astrophysics Postdoctoral Fellow}
\email{Joseph.Simon-1@colorado.edu}
\correspondingauthor{Maggie C. Huber}

\begin{abstract}
Supermassive black holes (SMBHs) and their host-galaxies coevolve through channels such as hierarchical merging and AGN feedback, establishing tight scaling relations between SMBH masses and properties of the host-galaxy. These scaling relations, particularly those with host-galaxy bulge mass (M${_\text{b}}$) and stellar velocity dispersion ($\sigma$), are widely used to estimate SMBH masses. Galaxy mergers drive gas into the centers of galaxies, enhancing SMBH growth and stellar bulge growth but not necessarily in tandem, so it is necessary to determine if galaxies undergoing a merger still follow SMBH-host-galaxy scaling relations. In this study, we explore scaling relations in galaxy mergers by taking advantage of the well-established value-added catalogs in the Sloan Digital Sky Survey. These catalogs provide close galaxy pairs, AGN broad lines for SMBH mass measurements, and bulge-disk decomposed stellar masses. We compare SMBH mass estimates from M$_\bullet-$M${_\text{b}}$ to SMBH mass estimates from AGN broad lines and find that SMBH mass growth outpaces the bulge for AGN residing in the secondary (less massive) bulge and for smaller physical separations between galaxies in a pair. Using our full sample of close galaxy pairs, we find that M$_\bullet-$M${_\text{b}}$ and M$_\bullet- \sigma$ predict significantly different fractions of major and minor black hole mergers with M$_\bullet-$M${_\text{b}}$ predicting black hole mass ratios closer to 1:1 and $6-20\%$ more major mergers overall. These results have major implications, including for predictions of astrophysical gravitational-waves and high-redshift overmassive SMBHs. 
\end{abstract} 

\section{Introduction}\label{introduction}

Supermassive black holes (SMBHs) and their host galaxies coevolve, as evidenced by scaling relations between the mass of the central SMBH (M$_\bullet$) and various properties of galaxy stellar mass, kinematics, and morphology \citep[e.g.,][]{Ferrarese2000,Gebhardt2000,Tremaine2002,Marconi2003,haring2004,Gultekin2009b,kh13,mm13,heckman2014,vandenbosch2016,graham2016,dn19}. Since direct SMBH mass measurements from stellar and/or gas dynamics in the SMBH's sphere of gravitational influence are only possible in the local universe (to a limiting distance of $\sim100$ Mpc), scaling relations are necessary to estimate SMBH mass for most galaxies. The most widely used scaling relations are with the velocity dispersion ($\sigma$) and stellar mass (M$_\text{b}$) of the galactic bulge. 

Both the M$_\bullet-$M$_\text{b}$ and M$_\bullet-\sigma$ relations appear in a vast array of astrophysical applications that require observationally-derived SMBH populations. Notable astrophysical studies that rely on scaling relations include modeling the SMBH binary population that is thought to produce the stochastic gravitational-wave background (GWB) recently detected by pulsar timing arrays (PTAs) \citep[e.g.,][]{agazie2023a,epta2024} and detectable by the upcoming Laser Interferometer Space Antenna (LISA) \citep[e.g,][]{colpi019,amaro-seoane2023}. LISA can also detect signals from individual massive black hole binaries (MBHBs), where  M$_\bullet-$M$_\text{b}$ is used in predictions of source populations \citep[e.g.,][]{izquierdo-villalba2023,drake2025}. Cosmological galaxy simulations are also calibrated to reproduce the observed local  M$_\bullet-$M$_\text{b}$ and M$_\bullet-\sigma$ scaling relations at $z=0$ for all galaxies \citep[e.g.,][]{eagle,tng,simba,Habouzit2021}, even at the low end of galaxy total stellar mass (M$_* \leq 10^{10.5}$ M$_\odot$). High-redshift AGN hosts found with JWST use local scaling relations to understand the SMBH population at earlier cosmic epochs. Many of these AGN show unusually overmassive SMBHs compared to the local M$_\bullet-$M$_\text{b}$, but are overall consistent with the local M$_\bullet-\sigma$ scatter \citep[e.g.,][]{maiolino2024,juodzbalis2026}.

Black hole-host-galaxy scaling relations have been extensively studied in the general galaxy population, with growing evidence supporting M$_\bullet-\sigma$ as more fundamental than M$_\bullet-$M$_\text{b}$. Empirical studies show M$_\bullet-\sigma$ consistently emerges as the relation with the least intrinsic scatter \citep[e.g.,][]{wake2012,vandenbosch2016,dn19,marsden2020,shankar2025,Huber2025}, and theoretical explanations favor a causal M$_\bullet-\sigma$ arising from AGN feedback \citep[e.g.,][]{king2015} and a non-causal M$_\bullet-$M$_\text{b}$ built up from hierarchical mergers \citep[e.g.,][]{peng2007,jahnke2011}. The divergence between M$_\bullet-$M$_\text{b}$ and M$_\bullet-\sigma$ has also been demonstrated for galaxies that are at higher redshift, disk-dominated, and/or have low total stellar mass \citep[e.g.,][]{cayenne2023,Huber2025}. Studies on this divergence are very limited for interacting galaxies. Merging galaxies are especially interesting for testing scaling relations, since both the SMBH and stellar bulge are growing in mass, but it is unknown whether they grow in tandem or not.

Hierarchical merging of galaxies and SMBHs can establish log-linear scaling relations between SMBH mass and host-galaxy properties  \citep[e.g.,][]{peng2007,johansson2009}, with the local M$_\bullet-$M$_\text{b}$ achieving low intrinsic scatter as a statistical outcome of the accumulation of mergers over time \citep[e.g.,][]{hirschmann2010, jahnke2011, tanaka2026}. However, mergers in progress could potentially cause galaxies to drift off of scaling relations if the SMBH and host-galaxy grow out of phase with one another; for example, some observations have shown overmassive SMBHs in galaxy mergers \citep{medling2015}. High-resolution hydrodynamical simulations align with these observations, finding that SMBH growth outpaces the host-galaxy bulge for an initially equal-mass SMBH binary \citep{prieto2021}. 

Merger-induced mass growth through channels such as triggering of active galactic nuclei (AGN) and star formation, which have been shown in both simulations \citep[e.g.,][]{barnes1991,dimatteo2005,moreno2019,he2023,schechter2025} and observations \citep[e.g.,][]{ellison2008,ellison11,patton11,patton13,comerford2015,barrows2023,comerford2024}, could potentially cause this out-of-phase growth between the black hole and host-galaxy depending on the merger scenario. It is known that levels of both AGN triggering and star formation are different between the more massive (primary) and less massive (secondary) galaxies in the merger, and this also depends on the overall galaxy mass ratio \citep[e.g.,][]{capelo2015,comerford2015, davies2015,steinborn2016,fu2018,yang2019,stemo2021,volonteri2022,barrows2023,steffen2023}.

Furthermore, the mass ratio of binary SMBHs embedded in a circumbinary disk evolves independently of the host galaxies via preferential accretion onto the secondary SMBH \citep[e.g.,][]{bate2002,farris2014,gerosa2015,munoz2020}, which can significantly boost the amplitude of gravitational-wave emission by driving SMBH mass ratios closer to 1:1 \citep[e.g.,][]{siwek2020,comerford2025}.

To understand how these various mass growth channels during a merger affect the black hole mass scaling relations, one needs a reliable SMBH mass estimate that is applicable to a large enough sample of galaxies for statistical significance. Single-epoch virial SMBH masses of broad-line AGN are ideal, since they probe dynamics within the SMBH sphere of influence and can be used for any galaxy that hosts an AGN with broad lines. However, previous single-epoch virial mass prescriptions assume the M$_\bullet-\sigma$ relation. With the recent single-epoch SMBH mass prescriptions from \cite{woo2026} that rely on dynamical modeling of the AGN broad line region (BLR), it is now possible to test how individual SMBH masses in galaxies undergoing a merger are offset from the M$_\bullet-$M$_\text{b}$ relation. Furthermore, the close galaxy pairs in the Sloan Digital Sky Survey (SDSS) with stellar mass and velocity dispersion measurements can be used to investigate the broader implications of using either M$_\bullet-$M$_\text{b}$ or 
M$_\bullet-\sigma$ to estimate properties such as black hole mass ratio ($q_\bullet$) and gravitational-wave chirp mass ($\mathcal{M}$).

In this paper, we will use the newly updated single-epoch SMBH mass prescriptions and the value-added catalogs in SDSS to thoroughly investigate how scaling relations predict the SMBH masses of galaxy mergers in progress. We will calculate a merger vs. nonmerger offset in SMBH masses for the subset of galaxies that have companions and host broad-line AGN. The "merger" SMBH mass will be the fiducial single-epoch virial mass for the broad-line AGN in these galaxy pairs, and the "nonmerger" SMBH mass will come from a power-law relation between the single-epoch SMBH mass and the bulge mass for AGN hosts without companions. We calculate an offset between these two SMBH mass estimates and observe how it changes due to bulge mass ratio and physical pair separation. We will then calculate the SMBH mass ratio and chirp mass for the entire sample of close galaxy pairs using M$_\bullet-$M$_\text{b}$ and
M$_\bullet-\sigma$ and quantify the divergence between their estimates. We will also test how our results are affected by indirectly inferring M$_\text{b}$ and $\sigma$, which is necessary when spectroscopy and bulge-disk decomposed photometry are not available.

In Section \ref{methods}, we will describe the SDSS sample used in this study, the power-law relation between single-epoch SMBH mass and bulge stellar mass for broad-line AGN hosts, and the calculations of $q_\bullet$ and $\mathcal{M}$. In Section \ref{results}, we show the results for the SMBH mass predicted by M$_\bullet-$M$_\text{b}$ vs. the mass estimated from AGN broad lines, the difference in $q_\bullet$ and $\mathcal{M}$ for the full close-pair sample, and the results when using an inferred bulge stellar mass and an inferred velocity dispersion. In Section \ref{discussion}, we interpret our results in terms of the physical processes during a galaxy merger that could influence the divergence between the true SMBH mass and the estimates from scaling relations. We expound on the implications for the astrophysical gravitational-waves in both the PTA and LISA frequency bands. Finally, in Section \ref{conclusion}, we summarize the key takeaways from our findings. 

Throughout this paper, we assume a flat cosmology with $\Omega_\Lambda=0.7$, $\Omega_\mathrm{M}=0.3$, and $H_0=70$ km s$^{-1}$ Mpc$^{-1}$.

\section{Methods} \label{methods}
\subsection{Data} \label{data}

\begin{figure}
    \centering
    \includegraphics[width=0.5\textwidth]{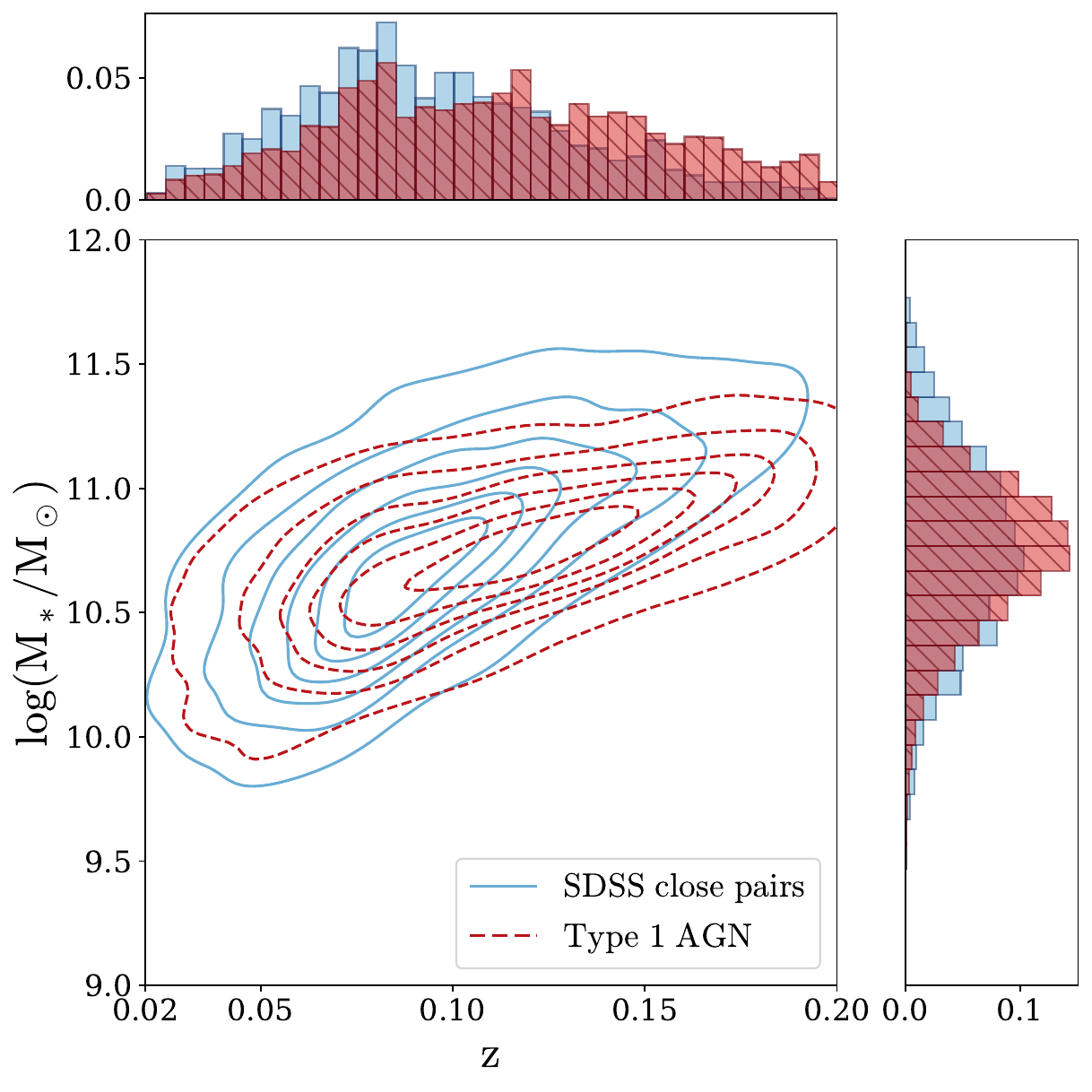}
    \caption{Total galaxy stellar mass (M$_*$) vs. redshift distribution for the galaxy close-pairs in light blue solid contours and histograms (Section \ref{section:closepairs}) and Type-1 broad line AGN hosts in dark red dashed contours and hatched histograms (\cite{liu19}, Section \ref{section:liu19}). This plot shows only the galaxies from each parent sample that pass the cuts described in Section \ref{methods} (see Tables \ref{tab:agnsample} and \ref{tab:closepairsample}). The contours show the joint $\log(\text{M}_*/\text{M}_\odot)$-$z$ distribution, the top panel shows the $z$ distribution, and the right panel shows the  $\log(\text{M}_*/\text{M}_\odot)$ distribution. From this figure, we can see substantial overlap between the total mass and redshift ranges of both samples, with the close-pairs shifted toward lower redshifts and extending over a larger total mass range than the Type 1 AGN.\label{fig:totalmassredshift}}
\end{figure}

The galaxy sample used in this project comes from SDSS DR7 \citep{sdssdr7} due 
to the availability of value-added catalogs that provide photometric bulge-
disk decompositions \citep{simard11,mendel13} and AGN broad-line detections 
\citep{liu19}. 

The complete DR7 spectroscopic galaxy database includes three different target selection algorithms corresponding to the main galaxy sample \citep{maingalaxy}, luminous red galaxies (LRGs) \citep{lrg}, and quasars \citep{qso}. The main galaxy sample has an extinction-corrected r-band Petrosian magnitude ($m_r$) limit of $m_r = 17.77$ and an r-band surface brightness ($\mu_{50}$) limit of $\mu_{50} < 24.5$ mag arcsec$^{-2}$. Since we require bulge masses for all results presented in this paper, all galaxies coincide with the \cite{mendel13} catalog, which uses the main galaxy sample exclusively.

We build our galaxy sample by cross-matching existing catalogs of AGN broad-line measurements \citep{liu19} and bulge stellar masses \citep{mendel13}, and identifying the galaxy close-pairs. In the following sections, we will describe how we combined these catalogs and derived quantities that we use for our analysis.

\subsubsection{Bulge+disk-decomposed stellar masses}\label{bd decomp}
Bulge-disk decomposed stellar mass and photometry for our sample come from the 
\cite{mendel13} and \cite{simard11} catalogs respectively. The \cite{simard11} catalog starts with the SDSS main galaxy sample, making additional surface brightness and magnitude cuts. They select galaxies with $14 \leq m_r \leq 17.77$ and $\mu_{50} \leq 23$ mag arcsec$^{-2}$ to create a spectroscopic galaxy sample with a lower magnitude limit to avoid brightness saturation.

We create a mass-complete sample from these catalogs by restricting the redshift to $0.02 \leq z \leq 0.2$ \citep{thanjavur2016}. We find that stellar masses are significantly affected by incompleteness at $\leq 10^{8.9}$ M$_{\odot}$ and make the mass cut accordingly.

For each galaxy, \cite{mendel13} reports total mass from both the one and two-
component fits to the surface brightness profile. The one-component total mass 
uses the mass inferred by fitting the entire galaxy, and the two-component fit 
uses the separate bulge and disk component masses added together. Following 
the advice from \cite{mendel13}, we exclude galaxies with disagreement between 
the total stellar mass from the one-component fit, M$_\mathrm{bulge+disk}$, 
and the mass from the two-component fit, M$_\text{bulge}+$M$_\mathrm{disk}$. 
\cite{mendel13} defines the offset between the total stellar masses as $
\Delta_{\mathrm{bulge+disk}}$ in units of standard error. We restrict our 
sample to galaxies where $\Delta_{\mathrm{bulge+disk}} < 1\sigma$, which  
removes unreliable masses from our sample. The minimum bulge-to-total stellar mass fraction of our sample is 0.01.

\subsubsection{Galaxy close-pairs}\label{section:closepairs}
In order to study the SMBH mass in SDSS close-pairs, we start by reproducing 
the spectroscopic close galaxy pair sample established in \cite{patton11}. We create a mass complete sample of SDSS 
DR7 galaxies by starting with the \cite{mendel13} catalog with the cuts described in Section \ref{bd decomp}. Close pairs are selected as galaxies that have a line-of-sight rest-frame
velocity separation $\Delta v < 500\mathrm{\ km \ s}^{-1}$, projected spatial 
separation $5 < r_p\mathrm{\ (kpc)} < 100$, and a galaxy stellar mass ratio 
$0.1 < q_\text{gal} < 10$. 

After we select our sample, we use $r_p$, $\Delta v$, $q_\text{gal}$ and $z$ to calculate the probability of a close-pair being a merger. We use the merger probability ($W$) from \cite{oleary2021}, where 

\begin{subequations}
    \label{mergerprob}
    \begin{equation}
    W(r_p,\Delta v, z)  = \frac{\exp(br_p)}{1+\exp[c_0(\Delta v-a)]}, 
    \end{equation}
    \begin{equation}
    a  = a_0(1+z)^{a_{z}} + a_r r_p, 
    \end{equation}
    \begin{equation}
    b  =  b_0 + (1+z)b_z.
    \end{equation}
\end{subequations}

We choose values of $a_0$, $a_z$, $a_r$, $b_0$, $b_z$, and $c_0$ according to the best-fit parameters from \cite{oleary2021} for major and minor mergers given a primary total stellar mass. For galaxies in multiple close-pairs, we keep the pair with the highest merger probability.

  \subsubsection{Type 1 AGN hosts}\label{section:liu19}
To calculate single-epoch virial SMBH masses and test scaling relations in merging galaxies, we need a sample of Type 1 AGN with broad lines and measurements of their widths. For our sample of SDSS DR7 galaxies that host Type 1 AGN, we use the catalog 
from \cite{liu19} which selects for galaxies with measured H$\alpha$ broad lines. These galaxies are selected from the entire SDSS DR7 dataset at $z < 0.35$ with either "Galaxy" or "Quasar" spectroscopic classifications. 

We use the H$\beta$ single epoch virial estimator to calculate single-epoch SMBH masses for these galaxies, since \cite{shen24} show that it correlates best with the SMBH 
masses from reverberation mapping. There are 379 Type 1 AGN that 
lack broad-line H$\beta$ measurements but have broad-line H$\alpha$, so we use 
the correlation between the full-width half maximum (FWHM) of H$\alpha$ and H$
\beta$ from \cite{greeneho2005} and the H$\beta$ single-epoch virial mass estimator of SMBH mass for these galaxies.

\subsubsection{Sample selection}

We cross-match the bulge-disk decomposition, close-pairs, and Type 1 AGN samples and apply cuts to create different samples to use in our analysis. Firstly, we create a sample of Type 1 AGN to fit a power-law between single-epoch virial SMBH mass (M$_\text{SE}$) and M$_\text{b}$. We apply the cuts on $z$, M$_*$, M$_\text{b}$ and $\Delta_{\mathrm{bulge+disk}}$ as described in Section \ref{bd decomp} and select galaxies with measured H$\alpha$ or H$\beta$ widths. We arrive at a sample of 2,315 galaxies to compare to other published scaling relations that do not make a distinction between mergers and nonmergers (Figure \ref{fig:mmbulgecomparison}, row 1 of Table \ref{tab:powerlaw}). To fit the M$_\text{SE}-$M$_\text{b}$ relation that we use as a formula to calculate SMBH masses, we restrict our sample to the 2,219 galaxies that do not have a companion within the $\Delta v$ and $r_p$ ranges described in Section \ref{section:closepairs} and are assumed to be nonmergers. The total stellar masses and redshifts of our sample is shown in Figure \ref{fig:totalmassredshift}.

Next, we select for Type 1 AGN in mergers, cross-matching the 2,315 galaxies with broad lines and reliable stellar masses with the close-pairs sample (with cuts on $\Delta v$, $r_p$ and $q_\text{gal}$ from Section \ref{section:closepairs}). We retain 75 galaxies in total with these cuts. Of these 75 galaxies, there are 52 where both the companion and the AGN host survive the cuts on M$_*$, M$_\text{b}$ and $\Delta_{\mathrm{bulge+disk}}$ (Section \ref{bd decomp}) and we can reliably compute the bulge stellar mass ratio ($q_\text{b}$). We use the Type 1 AGN hosts with companions to study the relative mass growth between the SMBH and the bulge (Section \ref{MSE results}). We do not make a cut on merger probability to retain as many data points as possible, instead showing the probability on a color scale and using it to weigh all of our results.

Lastly, we construct our sample of close-pairs to compare $q_\bullet$ and $\mathcal{M}$ estimates from M$_\bullet-$M$_\text{b}$ and 
M$_\bullet-\sigma$ (Section \ref{closepairs_results}, Section \ref{infresults}). To create this sample, we start by applying the aforementioned cuts on $z$, M$_*$, M$_\text{b}$, $\Delta_{\mathrm{bulge+disk}}$, $\Delta v$, $r_p$, and $q_\text{gal}$. Because we are using M$_\bullet-\sigma$, we also restrict our $\sigma$ measurements to fall within SDSS spectroscopic resolution constraints where $70 \leq \sigma  \mathrm{\ (km \ s}^{-1})\leq 420 $. We ensure that both companions in each galaxy pair adhere to these cuts, and then make a $W \geq 0.5$ cut on merger probability. We choose to apply the merger probability cut to this sample since the cuts still leave a statistically large sample of 591 close galaxy pairs.

A preview of our selected sample of broad-line AGN hosts can be found in Table \ref{tab:agnsample} and our sample of galaxies in close galaxy pairs are previewed in Table \ref{tab:closepairsample}, with machine-readable tables available for the entire dataset. Table \ref{tab:agnsample} provides all the data used to create Figures \ref{fig:mse_mbulge} through \ref{fig:other_correlations}, and Table \ref{tab:closepairsample} provides the data used to create Figures \ref{fig:qbh_distributions} through \ref{fig:qbh_inf_distributions}.

\begin{splitdeluxetable*}{cccccccBccc}
\tablecaption{
\label{tab:agnsample} Catalog of SDSS broad-line AGN host galaxies}
\tablehead{\colhead{objID}             & \colhead{RA} & \colhead{DEC}                  & \colhead{$z$} & \colhead{$\log\text{M}_\text{b}$}                                                                          & \colhead{$\sigma$}        & \colhead{$\log\text{M}_*$}         & \colhead{$\text{FWHM}_{\text{H}\beta}$} & \colhead{$\text{FWHM}_{\text{H}\alpha}$} & \colhead{$\log\text{L}_{5100}$}   \\ \colhead{}             & \colhead{(deg)} & \colhead{(deg)}                  & \colhead{} & \colhead{(M$_\odot$)}                                                                          & \colhead{(km s$^{-1}$)}        & \colhead{(M$_\odot$)}         & \colhead{(km s$^{-1}$)} & \colhead{(km s$^{-1}$)} & \colhead{(erg s$^{-1}$)}   }
\colnumbers
\startdata
588015509269708000 & 255.0571 & 40.1488 & 0.1069 & 10.236$^{+0.108}_{-0.151}$ & 151.7338 $\pm$ 12.0207 & 10.745 $\pm$ 0.1139  & 2262.4884 $\pm$ 31.8026   & 2262.4884 $\pm$ 156.9569 & 43.7349 $\pm$ 0.045  \\
587730775499735000 & 209.8668 & 1.3686  & 0.1175 & 9.336$^{+0.204}_{-0.243}$  & 74.6563 $\pm$ 17.8178  & 10.2804 $\pm$ 0.1267 & 1101.8289 $\pm$ 7.6663    & 1101.8289 $\pm$ 11.3602  & 42.7081 $\pm$ 0.0064 \\
588015509269905000 & 212.779  & 2.3136  & 0.1076 & 10.511$^{+0.081}_{-0.126}$ & 149.0361 $\pm$ 10.9112 & 10.6698 $\pm$ 0.083  & 5392.8522 $\pm$ 1292.8633 & 5392.8522 $\pm$ 257.8386 & 43.3777 $\pm$ 0.0501 \\
\enddata
\tablecomments{Columns: (1) SDSS DR7 object identification number, (2) right ascension in J2000.0 decimal degrees, (3) declination in J2000.0 decimal degrees, (4) spectroscopic redshift from SDSS, (5) bulge stellar mass from \cite{mendel13}, (6) aperture-corrected spectroscopic velocity dispersion from SDSS, (7) galaxy total stellar mass from \cite{mendel13}, (8) FWHM of broad H$\beta$ line from \cite{liu19}, (9) FWHM of broad H$\alpha$ line from \cite{liu19}, and (10) monochromatic luminosity at 5100 $\AA{}$ from \cite{liu19}. Each entry in the table has been reduced to two decimal places for readability in the preview, while all significant figures are provided in the machine readable version.}
Only a portion of this table is shown here to demonstrate its form and content. A machine-readable version of the full table is available.
\end{splitdeluxetable*}

\begin{splitdeluxetable*}{cccccBcccccBccccc}
\tablecaption{
\label{tab:closepairsample} Catalog of SDSS galaxies in close galaxy pairs}
\tablehead{\colhead{objID}   & \colhead{Companion ID}          & \colhead{RA} & \colhead{DEC}                  & \colhead{$z$} & \colhead{$\log\text{M}_\text{b}$}                                                                          & \colhead{$\sigma$}        & \colhead{$\log\text{M}_*$}         & \colhead{$\text{R}_\text{e}$} & \colhead{$n$} & \colhead{$g$} & \colhead{$r$} & \colhead{W} & \colhead{$\Delta v$} & \colhead{$r_p$} \\ \colhead{} & \colhead{}             & \colhead{(deg)} & \colhead{(deg)}                  & \colhead{} & \colhead{(M$_\odot$)}                                                                          & \colhead{(km s$^{-1}$)}        & \colhead{(M$_\odot$)}         & \colhead{(kpc)} & \colhead{} & \colhead{(mag)} & \colhead{(mag)} & \colhead{} & \colhead{(km s$^{-1}$)} & \colhead{(kpc)} }
\colnumbers
\startdata
587722981748113000 & 587722981748113000 & 156.0948 & 2.6289  & 0.1094 & 10.798$^{+0.048}_{-0.075}$ & 214.9298 $\pm$ 10.7232 & 11.0432 $\pm$ 0.0478 & 6.5048 $\pm$ 0.0784 & 6.92 $\pm$ 0.25 & 17.5  & 16.48 & 0.6549 & 90.5272  & 10.6848 \\
587722981748113000 & 587722981748113000 & 156.0955 & 2.6333  & 0.1091 & 9.379$^{+0.195}_{-0.226}$  & 126.1986 $\pm$ 13.4321 & 10.7222 $\pm$ 0.0926 & 3.1715 $\pm$ 0.0373 & 1.07 $\pm$ 0.05 & 17.75 & 17.1  & 0.6551 & 90.5546  & 10.6562 \\
587722982815694000 & 587722982815694000 & 198.367  & -1.2112 & 0.0719 & 10.177$^{+0.091}_{-0.135}$ & 133.2516 $\pm$ 9.6776  & 10.4571 $\pm$ 0.086  & 6.6456 $\pm$ 0.0831 & 6.46 $\pm$ 0.39 & 17.37 & 16.45 & 0.544  & 167.8168 & 10.6696 \\
\enddata
\tablecomments{Columns: (1) SDSS DR7 object identification number, (2) SDSS objID of galaxy companion, (3) right ascension in J2000.0 decimal degrees, (4) declination in J2000.0 decimal degrees, (5) spectroscopic redshift from SDSS, (6) bulge stellar mass from \cite{mendel13}, (7) aperture-corrected spectroscopic velocity dispersion from SDSS, (8) galaxy total stellar mass from \cite{mendel13}, (9) galaxy effective radius from \cite{simard11}, (10) galaxy S$\mathrm{\acute{e}}$rsic index from \cite{simard11}, (11) pure S$\mathrm{\acute{e}}$rsic model g-band magnitude from \cite{simard11}, (12) pure S$\mathrm{\acute{e}}$rsic model r-band magnitude from \cite{simard11}, (13) merger probability (Equation \ref{mergerprob}), (14) line-of-sight rest-frame velocity difference, and (15) projected physical separation between the close galaxy pair. Each entry in the table has been reduced to two decimal places for readability in the preview, while all significant figures are provided in the machine readable version.}
Only a portion of this table is shown here to demonstrate its form and content. A machine-readable version of the full table is available.
\end{splitdeluxetable*}

\subsection{Single-epoch virial mass estimates}
In this study, we use the recently updated single-epoch virial SMBH mass 
estimators from \cite{woo2026} as our fiducial SMBH masses. This estimator is 
based on reverberation mapping from 157 broad-line AGNs with reliable H$\beta$ time lags. These AGN have measured 
SMBH masses from the velocity dispersion ($
\sigma_\text{BLR}$) of the BLR and BLR size ($\text{R}_\text{BLR}$), where

\begin{equation}
    \text{M}_\bullet = f\frac{\sigma_\text{BLR}^2\text{R}_\text{BLR}}{\text{G}},
\end{equation}
where $f$ is the virial factor that encodes assumptions about the unknown 
geometry of the BLR. 

We use the virial factor from \cite{wang2026}, which measured $f$ without assuming a black hole-
host-galaxy scaling relation, and relies instead on dynamical modeling of the 
BLR for 38 individual quasars. The average virial factor they find (using the mean FWHM of the H$\beta$ broad-line) is $ \left< 
\log f \right> = -0.08\pm 0.06$. 

The single-epoch mass estimators from \cite{woo2026} incorporate a three-parameter fit between radius, luminosity, and Eddington ratio to account for the dependence of BLR size on the Eddington ratio. This allows for a more accurate single-epoch mass, where previous methods over-estimate the SMBH mass up to 0.5 dex. Since the relation from \cite{woo2026} uses $f$ based on M$_\bullet-\sigma$, we insert $f$ from \cite{wang2026} in the correction factor (C$_\text{M}$, see Section 4.3 of \cite{woo2026} ). Combining this updated radius-luminosity-Eddington ratio relation with the SMBH masses from reverberation mapping yields a single-epoch virial SMBH mass estimate where

\begin{multline}
\label{mse}
\log \left( \frac{\text{M}_{\mathrm{SE}}}{\text{M}_{\odot}} \right) = 
(6.24 \pm 0.26) 
\\
+ (0.48 \pm 0.04) \log \left(\frac{L_{5100,\mathrm{AGN}}}{10^{44}\mathrm{ \ erg \ s}
^{-1}}\right)
\\
+ (2.68 \pm 0.11)\left(\frac{\mathrm{FWHM_{H\beta}}}{1000 \mathrm{\ km \ s}
^{-1}}\right),
\end{multline}
with an intrinsic scatter of $0.21 \pm 0.02$ dex, where $L_{5100,\text{AGN}}$ is the monochromatic AGN
luminosity at $5100$\r{A} and  $\mathrm{FWHM_{H\beta}}$
is the 
full-width half maximum of broad-line H$\beta$. 
Equation \ref{mse} provides the ground-truth SMBH mass estimate for this study. Since this single-epoch virial prescription does not require an assumption of the M$_\bullet-\sigma$ relation, we treat it as an independent mass estimate to compare to the SMBH mass predicted by M$_\text{b}$.

For galaxies that only have measured H$\alpha$, we use the correlation from 
\cite{greeneho2005} where
\begin{multline}
\label{ha_hb}
\mathrm{FWHM_{H\beta}} = (1.07\pm0.07)
\\
\times10^3\left(\frac{\mathrm{FWHM_{H\alpha}}}{10^3\mathrm{\ km \ s}
^{-1}}\right)^{(1.03\pm0.03)}\mathrm{\ km \ s}^{-1},
\end{multline}
with an intrinsic scatter of 0.1 dex. We use this as the FWHM in Equation \ref{mse} to calculate single-epoch black 
hole mass for Type 1 AGN that only have H$\alpha$ measurements, propagating errors accordingly.

Equation \ref{mse} provides our fiducial SMBH mass to compare to M$_\bullet-$M$_\text{b}$ scaling relations, since the virial factor $f$ is independently derived and not based on an assumption about M$_\bullet-\sigma$. Although it depends on other assumptions about the BLR geometry, this estimate is a stronger inference of SMBH mass than host-galaxy scaling relations since it includes kinematic measurements closer to the black hole sphere of influence. Therefore, we can use it as a baseline to understand how the SMBH masses in galaxy mergers diverge from inferences based on stellar bulge masses.

\subsection{M$_\text{SE}-$M$_\text{b}$ relation for non-merging AGN hosts}

\begin{figure*}[ht]
    \centering
    \includegraphics[width=\textwidth]{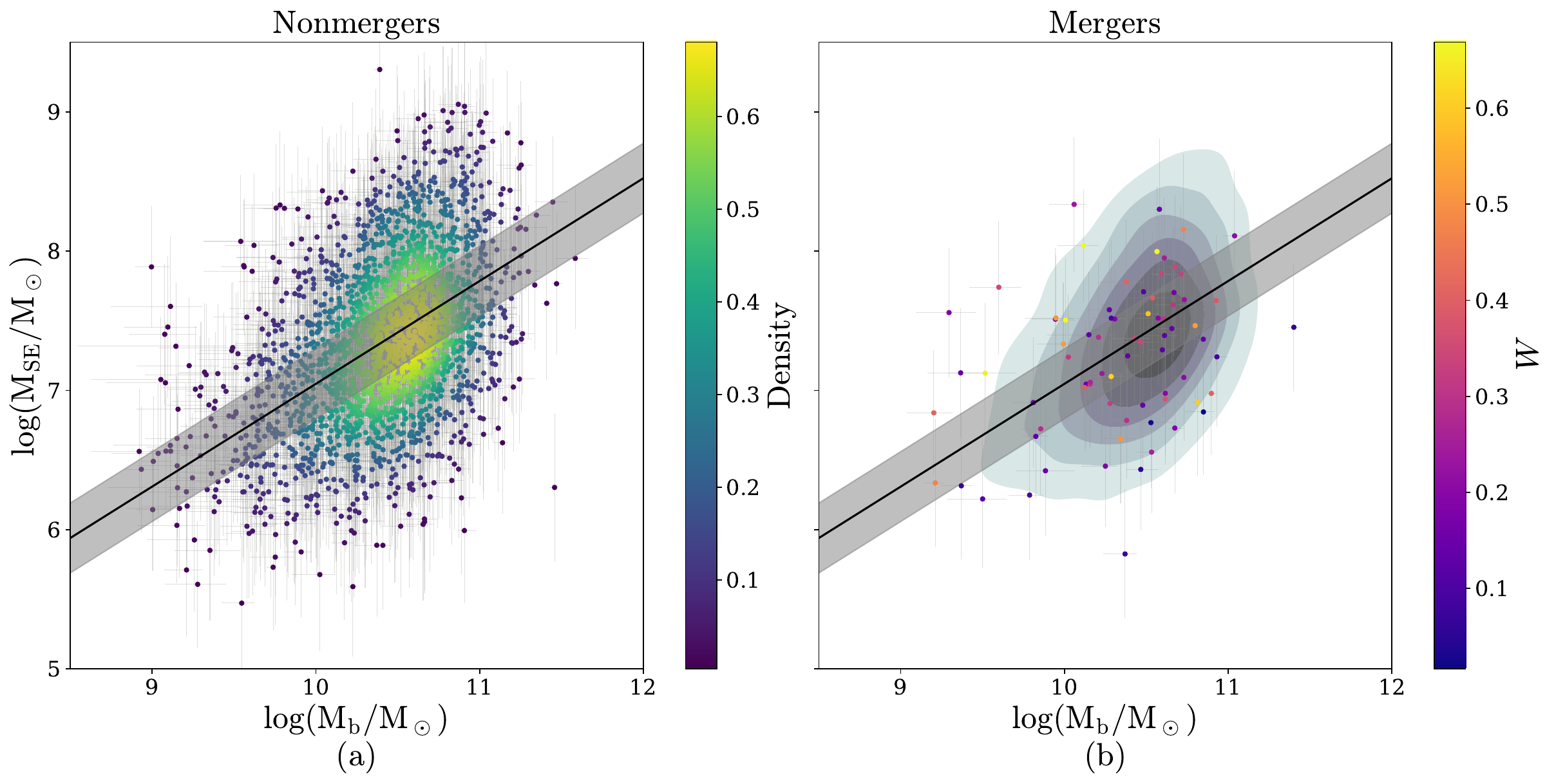}
    \caption{ \label{fig:mse_mbulge} Plots showing the best-fit power-law scaling relations between M$_\text{SE}$ (Equation \ref{mse}) and M$_\text{b}$ (Table \ref{tab:powerlaw}). Figure (a) shows the 2,219 non-merging Type 1 AGN hosts. The black line is the best-fit scaling relation (based on the nonmerger-only scaling relation in the second row of Table \ref{tab:powerlaw}) and the color bar is the Gaussian kernel density estimator showing the concentration of points. The gray shaded region is the intrinsic scatter around the scaling relation. Figure (b) has the 2,219 nonmergers as a 2D contour plot in the background, with the same scaling relation as in Figure (a) and 75 Type 1 AGN hosts in a close-pair in the foreground. The color bar refers to the merger probability of the close-pair to which the AGN belongs (Equation \ref{mergerprob}).}
\end{figure*}

\begin{figure}
    \centering
    \includegraphics[width=0.5\textwidth]{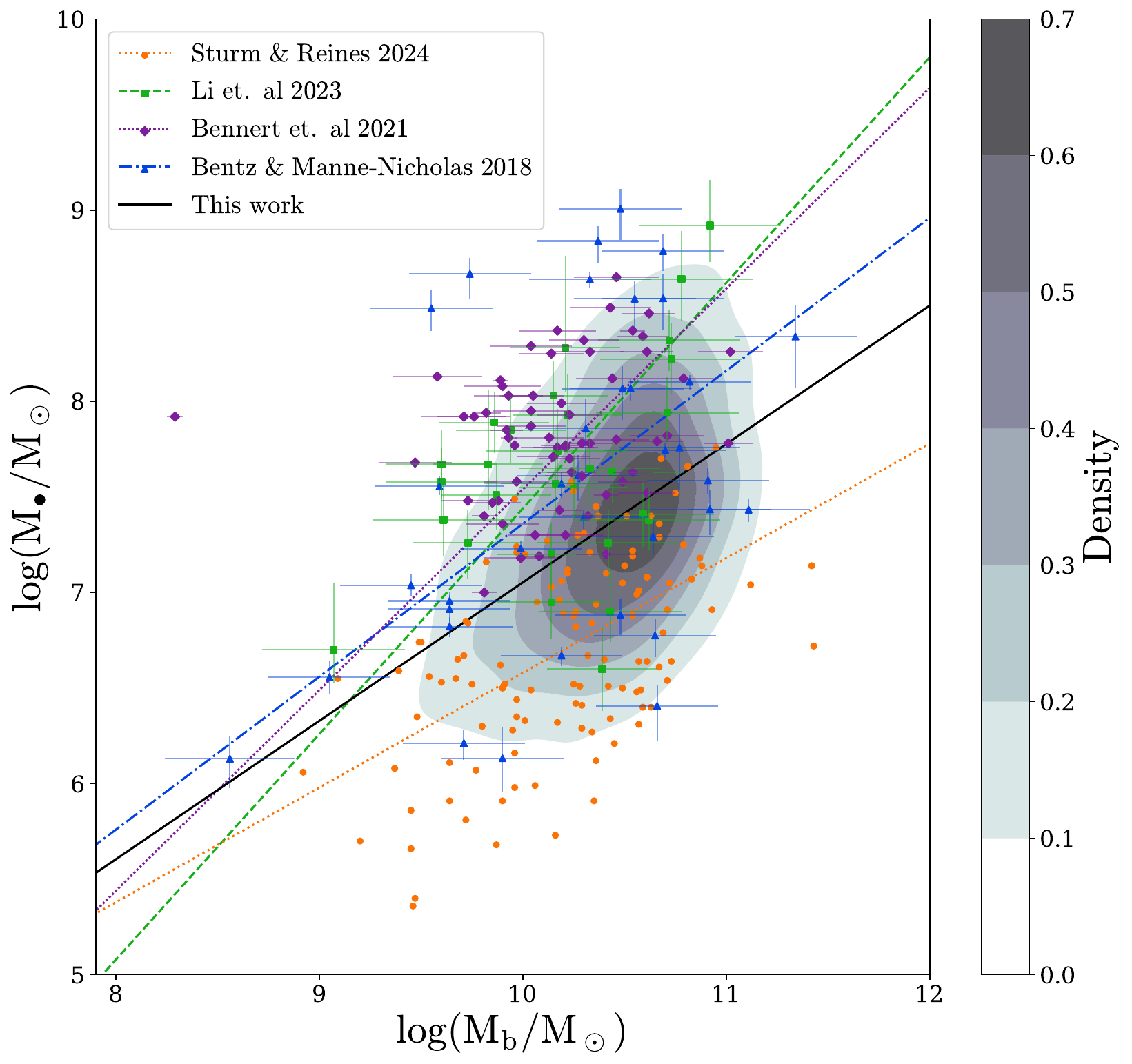}
    \caption{Comparison between our best-fit power-law scaling relation between M$_\text{SE}$ and M$_\text{b}$ (mergers+nonmergers, first row of Table \ref{tab:powerlaw}) and previously published M$_\bullet-$M$_\text{b}$ scaling relations from AGN broad lines and photometric bulge mass. Our sample of 2,219 Type 1 AGN are plotted as 2D contours corresponding to the Gaussian kernel density of points and our scaling relation is the black line. The data points and scaling relations of \cite{bentz18} (blue triangles), \cite{bennert2021} (purple diamonds), \cite{li2023} (green squares), and \cite{sturm24} (orange circles) are shown for comparison. Descriptions of each sample and scaling relation parameters are in Table \ref{tab:powerlaw}. The SMBH masses in this work are less massive relative to all published scaling relations aside from \cite{sturm24}. \label{fig:mmbulgecomparison}}
\end{figure}

\begin{splitdeluxetable*}{ccccBcccc}
\tablecaption{
\label{tab:powerlaw} M$_\bullet$-M$_\text{b}$ for Type 1 AGN hosts
($\log_{10}(\text{M}_{\bullet}/\text{M}_{\odot}) = \alpha + \beta \log_{10}($M$_\text{b}/10^{11}$M$_{\odot})$)}
\tablehead{\colhead{Relation}             & \colhead{Sample size} & \colhead{$z$}                  & \colhead{$\log_{10}$(M$_{\bullet}$/M$_\odot$) range} & \colhead{M$_{\bullet}$}                                                                          & \colhead{$\alpha$}        & \colhead{$\beta$}         & \colhead{$\epsilon$}   }
\startdata
This work (all)           & 2,315      & 0.02 $\leq$ $z$ $\leq$ 0.2     &  5.45 - 9.32   & SE H$\beta$ $\text{\citep{woo2026}}$                                   & 7.78$\pm$0.02 & 0.72$\pm$0.03 & 0.25$\pm$0.02 \\
This work (nonmergers)            & 2,219      & 0.02 $\leq$ $z$ $\leq$ 0.2     & 5.45 - 9.32   & SE H$\beta$ $\text{\citep{woo2026}}$                              & 7.79$\pm$0.02 & 0.74$\pm$0.03 & 0.25$\pm$0.02 \\
$\text{\cite{sturm24}}$ & 117         & 0 $<$ $z$ $\leq$ 0.055  & 5.36 - 7.76  & SE H$\alpha$ $\text{\citep{reines13}}$ &  7.18$\pm$0.09 & 0.60$\pm$0.11 & n/a     \\ 
$\text{\cite{li2023}}$ & 38 & 0.02 $\leq$ $z$ $\leq$ 0.8 & 6.60 - 8.92 & RM $\text{\citep{grier2017}}$ & $8.62^{+0.89}_{-0.68}$ & $1.18^{+0.76}_{-0.52}$ & 0.39$\pm$0.11 \\
$\text{\cite{bennert2021}}$ & 63 & 0.02 $\leq$ $z$ $\leq$ 0.1 &  7.0 - 8.65 & SE H$\beta$ $\text{\citep{bennert2015}}$  & $8.59\pm0.06$ & $1.05 \pm 0.07$ & $0.12 \pm 0.1$ \\
$\text{\cite{bentz18}}$     & 37          & 0 $<$ $z$ $<$ 0.15        &  6.13 - 9.01  & RM $\text{\citep{bentz15}}$                          & 8.16$\pm$0.45 & 0.80$\pm$0.30 & 0.55$\pm0.16$ \\
\enddata
\tablecomments{SE refers to the single-epoch virial SMBH mass estimate and RM refers to reverberation mapping. The intrinsic scatter parameter is in the last column as $\epsilon$. These scaling relations are shown for comparison in Figure \ref{fig:mmbulgecomparison}.}
\end{splitdeluxetable*}

Since we are using $\log($M$_\text{SE}/$M$_\odot)$ as our fiducial SMBH 
mass, we want to ensure that we are also using it in our black hole-host 
galaxy scaling relation. To do this, we use our sample of 2,219 galaxies with broad-line AGN that are not in our close-pair catalog (we refer to these galaxies as nonmergers), in order to calculate $\log($M$_\text{SE}/$M$_\odot)$ from Equation 
\ref{mse}. We then fit a power-law scaling relation between $\log($M$_\text{SE}/$M$_\odot)$ and $\log($M$_\text{b}/$M$_\odot)$ from \cite{mendel13}.

We do not fit a relation to M$_\bullet-\sigma$ for our AGN sample due to AGN contamination of the spectra. Since we do not have enough spatial resolution to de-blend the AGN source from the starlight continuum and measure accurate $\sigma$, the dispersion measurements in the SDSS catalog can be overestimated by up to a factor of 3 \citep{winkel2025}. Therefore, we opt to only fit the M$_\text{b}$ power-law scaling relation. This should not be an issue for our results in Sections \ref{closepairs_results} and \ref{infresults}, since the majority of our pairs do not host an AGN and we are testing established scaling relations instead of fitting a new M$_\bullet-\sigma$. 

Since our measured bulge masses do not account for possible contamination from AGN luminosity, we examine the possibility that we overestimate the bulge mass of some AGN hosts in our sample. \cite{sturm24} recently explored this issue for a sample of broad-line AGN in SDSS,
where they refit the galaxy surface brightness profiles including an AGN 
component in addition to the bulge and disk. For their sample of SDSS broad-line
AGN, \cite{sturm24} found that their AGN-subtracted bulge masses were consistent with those from \cite{mendel13} with a median offset of $0.03 \pm 0.20$ dex. Since our sample of broad-line
AGN cover the same luminosity range as that of \cite{sturm24}, we use
\cite{mendel13} for our galaxy mass measurements because it is a widely used and
tested catalog that has proven to be robust.

The 
scaling relation we use takes the form
\begin{equation}
    \log_{10}(\text{M}_\text{SE}/\text{M}_{\odot}) = \alpha + \beta\log_{10}\left(\frac{\mathrm{M}_\text{b}}{10^{11} \mathrm{ \ M}_\odot}\right),
\label{msescaling}
\end{equation}
Where $\alpha$ is the intercept and $\beta$ is the slope. We perform Bayesian linear regression with the \texttt{linmix} software package \citep{kelly07}, which incorporates the measurement errors in both variables and fits for intrinsic scatter. Figure \ref{fig:mse_mbulge} shows our power-law and the best-fit parameters are in the second row of Table \ref{tab:powerlaw}. We also fit a power-law to our full Type 1 AGN host sample, with both mergers and nonmergers (first row of Table \ref{tab:powerlaw}) to compare to previous work.

The remaining rows of Table \ref{tab:powerlaw} show other recently published M$_\bullet-$M$_\text{b}$ scaling relations using photometric bulge masses and black hole masses based on AGN broad lines. The different samples and scaling relations are shown for comparison in Figure \ref{fig:mmbulgecomparison}. The scaling relation we fit in this work has the largest number of galaxies, and is complete in total and bulge stellar mass at our redshift range. Our scaling relation is consistent with the range of published slope $\beta$ and intrinsic scatter $\epsilon$ values. We measure an intercept $\alpha$ slightly greater than \cite{sturm24} and slightly less than all other published intercepts. The majority of our single-epoch virial masses, like those from \cite{sturm24}, are found to be less massive compared to the other local M$_\bullet-$M$_\text{b}$ relations. However, this offset is not as extreme since our sample extends to SMBH masses about an order of magnitude greater than the upper limit of the \cite{sturm24} sample.

Our new scaling relation offers a way of estimating the SMBH mass for AGN hosts based on a sample with single-epoch virial masses ranging from $10^{5.5}$ M$_\odot \lesssim$ M$_\bullet \lesssim 10^{9.3}$ M$_\odot$. We use it to compare to the measured single-epoch virial SMBH masses of AGN in close-pairs, investigating how black hole masses in mergers are offset from predictions based on the bulge masses of non-merging galaxies. 

\subsection{M$_\bullet-$M$_\text{b}$ and M$_\bullet-\sigma$ for galaxy close-pairs} \label{pairs} 

\begin{deluxetable*}{clccc}
\tablecaption{Black hole mass scaling relations for galaxies in close-pairs in the form $\log_{10}(\text{M}_{\bullet}/\text{M}_{\odot}) = \alpha + \beta \log_{10}(X)$ with intrinsic scatter $\epsilon$. \label{tab:scalingrel}}
\tablehead{\colhead{Relation} & \colhead{X} & \colhead{$\alpha$}        & \colhead{$\beta$}         & \colhead{$\epsilon$}}
\startdata
\text{\cite{mm13}} & $\sigma$/200 km s$^{-1}$& $8.32 \pm 0.05$ & $5.64 \pm0.32$ & $0.38$          \\
            & M$_\text{b}$/10$^{11}$M$_\odot$ & $8.46 \pm 0.08$ & $1.05 \pm 0.11$ & $0.34$          \\
\cite{kh13} & $\sigma$/200 km s$^{-1}$               & $8.49 \pm 0.05$ & $4.38 \pm 0.29$ & $0.29$          \\
            & M$_\text{b}$/10$^{11}$M$_\odot$ & $8.69 \pm 0.05$ & $1.16 \pm 0.08$ & $0.29$          
\enddata

\end{deluxetable*}

We extend our study to our full sample of galaxy close-pairs, focusing on how known differences between commonly-used  M$_\bullet-$M$_\text{b}$ and M$_\bullet-\sigma$ scaling relations impact the calculation of the merger black hole mass ratio ($q_\bullet$) and chirp mass ($\mathcal{M}$).

The SMBH mass scaling relations that we consider here are from 
\cite{mm13} and \cite{kh13}. These scaling relations are widely used and span the range of parameter values that have been measured for M$_\bullet-$M$_\text{b}$ and M$_\bullet-\sigma$. The parameters of each scaling relation are given in Table \ref{tab:scalingrel}.  
 
 For the M$_\bullet-$M$_\text{b}$ relation, both \cite{mm13} and \cite{kh13} cover a similar dynamic range of SMBH mass and only include early-type, elliptical galaxies with classical bulges. They differ in M$_\bullet-\sigma$, as \cite{mm13} incorporates both late- and early-type galaxies in their fit and \cite{kh13} excludes any galaxies without a classical bulge from their M$_\bullet-\sigma$. \cite{mm13} covers a larger dynamic range of SMBH mass in M$_\bullet-\sigma$ as a result, with their sample going down to M$_\bullet \sim 10^6$ M$_\odot$ with around 10 galaxies with M$_\bullet < 10^7$ M$_\odot$. \cite{kh13} only has one galaxy with  M$_\bullet < 10^7$ M$_\odot$. This is a subtle difference, but as we investigate here, it can significantly impact the estimation of $q_\bullet$.

We calculate SMBH masses for our galaxy sample from \cite{mm13} and \cite{kh13} M$_\bullet-$M$_\text{b}$, using the M$_\text{b}$ from the \cite{mendel13} catalog and the inferred M$_\text{b}$ from the prescription in Section \ref{infbulge}. 

For SMBH mass from M$_\bullet-\sigma$, we use both spectroscopic and inferred velocity dispersions (Section \ref{infbulge}). For spectroscopic velocity dispersion, we obtain spectra for our galaxies from SDSS and perform an aperture correction \citep{bezanson2011} where 
\begin{equation} \label{apcorr}
    \sigma = \sigma_{ap}(8.0r_{ap}/R_e)^{0.066}.
\end{equation}
Here, $\sigma_{ap}$ is the velocity dispersion from SDSS, $r_{ap} = 1.5 \arcsec $ is the spectroscopic fiber radius, and $R_e$ is the effective radius from the half-light radius and ellipticity given in \cite{simard11}. 

Using M$_\text{b}$ and $\sigma$, we go on to calculate SMBH masses from scaling relations for all galaxies in our sample. We use a Monte Carlo approach to error propagation to create probability distributions for all measurements and scaling relation parameters and take 700 random samples from each distribution, where 700 draws is the threshold beyond which the median SMBH masses change by less than $1\%$ between repeat calculations. The result is 700 samples that represent the SMBH mass probability distribution. We use the median of this distribution as the measured value and standard deviation as the associated errors. 

\subsubsection{Inferred M$_\text{b}$ and $\sigma$} \label{infbulge}
In practice, bulge mass is often inferred when only single-component photometry is available. This scenario is commonly encountered in modeling the GWB \citep[e.g.,][]{sesana2013a,ravi2015,arzoumanian2021,casey-clyde2022,agazie2023a}, and there are a number of different prescriptions to estimate the bulge mass fraction ($f_\text{b}$). In this paper, we follow the improved $f_\text{b}$ prescription from \cite{Huber2025} with the color cut from \cite{bluck2014}, where the color
\begin{equation}
\label{color}
    (g-r) > 0.06 \times \log(\text{M}_*/\text{M}_{\odot}) - 0.01
\end{equation}
designates red/early-type galaxies, and others are considered blue/late-type galaxies. We follow the $f_\text{b}$ prescription derived in \cite{Huber2025} and randomly select $f_\text{b}$ from a uniform distribution that covers the interval $[\langle f_\text{b}\rangle - 0.1,\langle f_\text{b}\rangle + 0.1]$. For early-type galaxies
\begin{multline} \label{fb1}
\langle f_\text{b}\rangle = 
\begin{cases}
    0.9, \\
    0.25 + 0.325(\log_{10}(\text{M}_* / \text{M}_\odot) - 9), \\ 
    0.25, 
\end{cases} 
    \\
    \hspace{3.5cm}\text{M}_* (\text{M}_\odot) >  10^{11} \\
    \hspace{2.38cm}10^{9} < \text{M}_* (\text{M}_\odot ) \leq 10^{11} \\
    \hspace{3.5cm}\text{M}_* (\text{M}_\odot ) \leq 10^{9}.
    \\
\end{multline}

For late-type galaxies, 
\begin{multline}\label{fb2}
\langle f_\text{b}\rangle = 
\begin{cases}
    0.25, \\
    0.10 + 0.075(\log_{10}(\text{M}_* / \text{M}_\odot) - 9), \\ 
    0.10, 
\end{cases} 
    \\
    \hspace{3.5cm}\text{M}_* (\text{M}_\odot ) >  10^{11} \\
    \hspace{2.38cm}10^{9} < \text{M}_* (\text{M}_\odot ) \leq 10^{11}  \\
    \hspace{3.5cm}\text{M}_* (\text{M}_\odot ) \leq 10^{9} .
    \\
\end{multline}

We use inferred $f_\text{b}$ and multiply by the total stellar galaxy mass M$_*$ to calculate the inferred bulge mass M$_{\mathrm{b,inf}}$ for our full close-pairs sample.

An inferred velocity dispersion can also be readily calculated for our sample from the virial theorem. We use the inferred velocity dispersion from \cite{bezanson2011} that can be applied to all galaxy morphologies using a virial constant $K_{\nu}(n)$ that depends on S$\mathrm{\acute{e}}$rsic index $n$ where 
\begin{equation}
    K_\nu(n) = \frac{73.32}{10.465 + (n-0.94)^2} + 0.954.
    \label{virial constant}
\end{equation}
Inferred velocity dispersion can be calculated from $K_{\nu}(n)$, the total stellar mass M$_*$, and effective radius $R_e$ with 
\begin{equation}
    \sigma_{inf} = \sqrt{\frac{\text{GM}_{*}}{0.557K_{\nu}(n)R_e}}.
\label{inferred veldisp}
\end{equation}

Previously, \cite{Huber2025} showed that inferring $\sigma$ yields estimates much closer to the measured spectroscopic $\sigma$ than M$_{\mathrm{b,inf}}$ was able to achieve relative to M$_\text{b}$ from bulge-disk decomposition. Here, we will test how these inferences perform relative to each other in the calculation of $q_\bullet$ when we use inferred $\sigma$ and M$_\text{b}$ in the SMBH mass-host-galaxy scaling relations.

\subsubsection{SMBH mass ratio $q_\bullet$ and chirp mass $\mathcal{M}$}
The SMBH masses from scaling relations are used to estimate key SMBH binary properties. In this study, we investigate how differences in scaling relations impact the inferred SMBH mass ratio and the chirp mass.

We define the black hole mass ratio as 
\begin{equation}
    q_\bullet = \mathrm{M}_{\bullet,2}/\mathrm{M}_{\bullet,1},
    \label{q}
\end{equation}
Where $\mathrm{M}_{\bullet,1}$ is the primary SMBH mass and $\mathrm{M}_{\bullet,2}$ is the secondary ($\mathrm{M}_{\bullet,1} > \mathrm{M}_{\bullet,2}$). We distinguish major and minor SMBH pairs based on their mass ratio, where $q_\bullet \geq 0.25$ are major pairs and $q_\bullet < 0.25$ are minor.

The mass ratio $q_\bullet$ is also used to measure the chirp mass, the quantity that directly scales with gravitational-wave amplitude produced by a SMBH merger. Gravitational wave emission in both the PTA and LISA bands strongly depends on the distribution of SMBH binary chirp masses, which can be indirectly measured using SMBH mass scaling relations \citep{agazie2023a,drake2025}. Here, we will propagate scaling relations to individual $\mathcal{M}$ calculations for our close-pair sample to show how $\mathcal{M}$ diverges between M$_\bullet-$M$_\text{b}$ and M$_\bullet-\sigma$. $\mathcal{M}$ is defined as
\begin{equation}
    \label{chirpmass}
    \mathcal{M} = \left[ \frac{q_\bullet}{(1+q_\bullet)^2}\right]^{3/5}\mathrm{M}_{\bullet,\mathrm{tot}},
\end{equation}
where $q_\bullet$ is from Equation \ref{q} and $\mathrm{M}_{\bullet,\mathrm{tot}}$ is the total SMBH mass ($\mathrm{M}_{\bullet,\mathrm{tot}} = \mathrm{M}_{\bullet,1} + \mathrm{M}_{\bullet,2}$).

For each $q_\bullet$ estimate, we only plot the values of $q_\bullet > 0.01$ to ensure we are removing nonphysical SMBH mass ratios that may be dominated by measurement error in host-galaxy properties. Therefore, each plot of $q_\bullet$ will show a slightly different number of galaxies, though most sample sizes are consistent within $N \simeq 10$ galaxies and only the $q_\bullet$ from $\sigma_\text{inf}$ has a notably different sample size of $N \simeq 500$ (Figure \ref{fig:qbh_inf_distributions} (c) and (d)). 

\section{Results} \label{results}
\subsection{Comparing M$_\bullet$ measurements from $\text{M}_{SE}$ and $\text{M}_{SE}-\text{M}_\text{b}$ for AGN hosts in close-pairs} \label{MSE results}

\begin{figure*}[ht]
    \centering
    \includegraphics[width=0.9\textwidth]{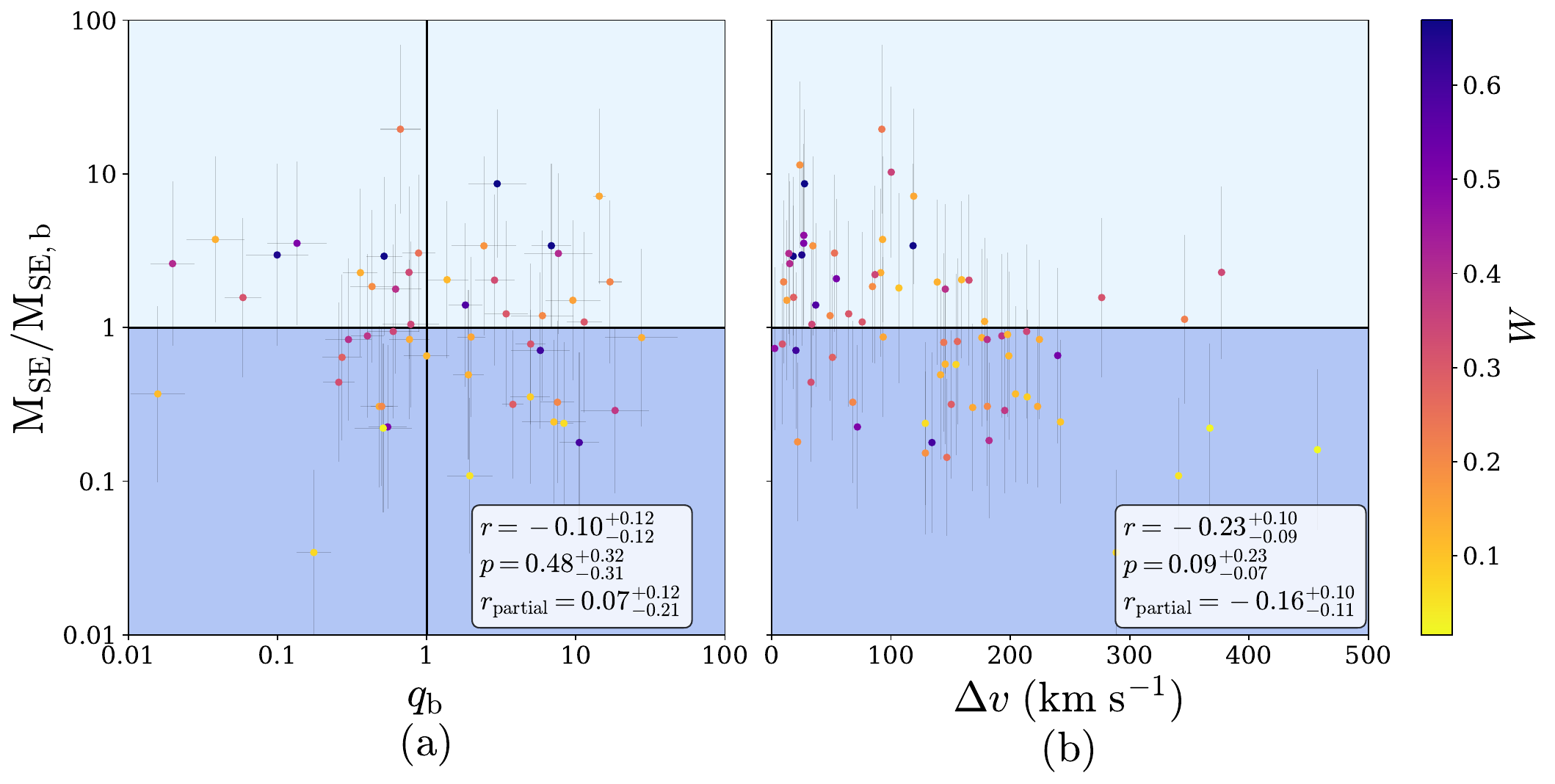}
    
    \hspace{-1.55cm}\includegraphics[width=0.8\textwidth]{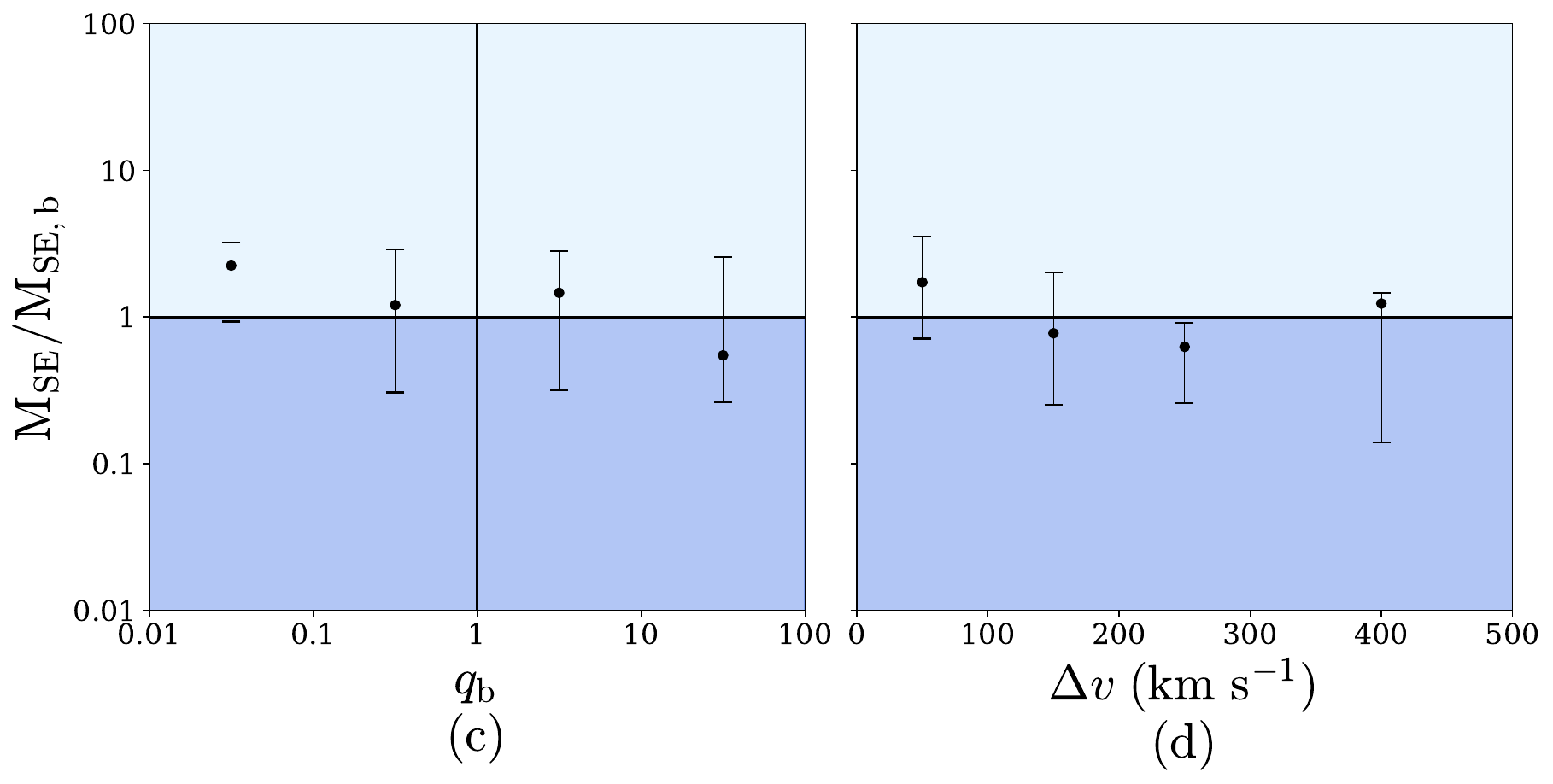}

    \caption{\label{fig:other_correlations} The ratio between the SMBH mass calculated from the single-epoch virial method M$_\text{SE}$ from Equation \ref{mse} and the SMBH mass M$_\text{SE,b}$ calculated from Equation \ref{msescaling} for AGN hosts in close-pairs vs. the galaxy bulge stellar mass ratio $q_\mathrm{b}$ and the velocity separation $\Delta v$ between the galaxies in a pair. The top panels (a) and (b) show each data point with a color scale to show galaxy merger probability from Equation \ref{mergerprob}. The Pearson's-r correlation coefficients are in the boxes in the bottom right corner, with the probability-weighted $r$, the associated p-value, and the partial correlation coefficient $r_\text{partial}$ with M$_\text{b, AGN host}$ as the confounding variable. The light blue and dark blue regions of the plot show where the single-epoch SMBH mass (our ground truth) is greater than and less than the SMBH mass estimate based on stellar bulge mass, respectively. The bottom panels (c) and (d) show the results binned by the x-axis and spaced for clarity, with the data points as the probability-weighted average offset and the error bars as the 16$^\text{th}$ and 84$^\text{th}$ percentiles as described in Table \ref{tab:offsets}. From this figure, we can see that overmassive SMBHs preferentially exist in secondary bulges in more minor mergers and more closely separated pairs.}
\end{figure*}

\begin{deluxetable*}{clcccc}
    \tablecaption{Average offsets (weighted by merger probability) between single-epoch virial SMBH mass and SMBH mass predicted from the nonmerger M$_\text{SE}-\mathrm{M}_\text{b}$ relation (M$_\text{SE}$/M$_\text{SE,b}$) binned by bulge mass ratio $q_\text{b}$ and velocity separation $\Delta v$ (data shown in Figures \ref{fig:other_correlations}(c) and (d)). \label{tab:offsets}}
    \tablehead{\colhead{Merger property} & \colhead{Range} & \colhead{N$_\mathrm{gal}$ } & \colhead{M$_\text{SE}$/M$_\text{SE,b}$} & \colhead{p$_{16}$}         & \colhead{p$_{84}$}}
    \startdata
$q_\mathrm{b}$ & $0.01 < q_\text{b} \leq 0.1$          & 5  & 2.30 & 1.04 & 3.17 \\
               & $0.1 < q_\text{b} \leq 1$             & 20 & 1.24 & 0.29 & 3.04 \\
               & $1 < q_\text{b} \leq 10$              & 21 & 1.49 & 0.31 & 2.65 \\
               & $10 < q_\text{b} \leq 100$            & 6  & 0.54 & 0.26 & 2.66 \\
$\Delta v$     & $0 < \Delta v$ (km s$^{-1})$ $ \leq 100$    & 33 & 1.75 & 0.70 & 3.64 \\
               & $100 < \Delta v$ (km s$^{-1})$ $ \leq 200$  & 28 & 0.78 & 0.24 & 2.06 \\
               & $200 < \Delta v$ (km s$^{-1})$ $ \leq 300$  & 9  & 0.68 & 0.27 & 0.93 \\
               & $300 < \Delta v$ (km s$^{-1})$ $ \leq 500$ & 5  & 1.18 & 0.14 & 1.34
    \enddata
    \tablecomments{N$_\text{gal}$ is the number of galaxies in the specified $q_\text{b}$ and $\Delta v$ ranges. p$_{16}$ and p$_{84}$ are the 16$^\text{th}$ and 84$^\text{th}$ percentiles of M$_\text{SE}$/M$_\text{SE,b}$ in the specified galaxy bins.}
\end{deluxetable*}

For our first analysis, we compare the SMBH masses from single-epoch virial estimates (Equation \ref{mse}) of broad-line AGN in galaxy mergers to the SMBH masses obtained from using M$_\text{b}$ in the nonmerger M$_\text{SE}-$M$_\text{b}$ relation from Equation \ref{msescaling} (M$_\text{SE,b}$). By comparing these two mass estimates, we can reveal whether the SMBH and bulge are growing together or if one is significantly outpacing the other.  This can show where scaling relations break down in galaxy mergers, especially if we can connect relevant merger properties to which pairs show the greatest difference in growth between the SMBH and bulge. 

To calculate the offset between scaling relations and single epoch virial estimates, we take the ratio M$_\text{SE}$/M$_\text{SE,b}$ and plot it as a function of various galaxy merger properties. We also calculate Pearson's-r correlation coefficients, using Monte Carlo sampling of 700 draws to propagate measurement errors and weighting the results by merger probability.

For any statistical test in our results, including the Pearson's-r correlation coefficients, we show corresponding p-values where our minimum sensitivity is $p_\text{min} = 0.001$ and p-values below this threshold are expressed as $p < p_\text{min}$.

We show our results for the bulge stellar mass ratio and the velocity separations; we tested the galaxy stellar mass ratio and the projected separation and found no evidence for correlation with M$_\text{SE}$/M$_\text{SE,b}$.

The results are shown in Figure \ref{fig:other_correlations} and Table \ref{tab:offsets}, which show evidence for slight negative correlations of $q_\text{b}$ and $\Delta v$ with the mass offset M$_\text{SE}$/M$_\text{SE,b}$. The results for $q_\text{b}$ are shown in Figures \ref{fig:other_correlations} (a) and (c). We specifically designate $q_\text{b}$ as the bulge mass ratio of the AGN host to the companion ($q_\text{b} = \text{M}_\text{b, AGN host} / \text{M}_\text{b, companion}$) such that data points left of the vertical line at $q_\text{b} = 1$ are where the AGN host resides in the secondary bulge and points to the right are where the AGN host resides in the primary bulge. We see that the Pearson's-r between $q_\text{b}$ and M$_\text{SE}$/M$_\text{SE,b}$ is -0.13, and that overmassive SMBHs by a factor of 2 on average are located in secondary bulges in more extreme bulge mass ratios. However, we also compute the partial correlation coefficient ($r_\text{partial}$) to account for the confounding variable of the bulge mass of the AGN host, on which both $q_\text{b}$ and M$_\text{SE}$/M$_\text{SE,b}$ depend. We use a simple calculation given the individual Pearson $r$ values between variables $x=q_\text{b}$ or $\Delta v$, $y=\log(\text{M}_\text{SE}/\text{M}_\text{SE,b}),$ and $z=\log(\text{M}_\text{b, AGN host}/\text{M}_\odot)$ where
\begin{equation}
    r_p = \frac{r_{xy}-r_{xz}r_{yz}}{\sqrt{1-r_{xz}^2}\sqrt{1-r_{yz}^2}}.
\end{equation}

The partial correlation coefficient with M$_\text{b}$ for $q_\text{b}$ is 0.01, indicating that the strong correlation of the bulge mass with both $q_\text{b}$ and M$_\text{SE}$/M$_\text{SE,b}$ is responsible for the slight negative trend.

Figure \ref{fig:other_correlations} (b) and (d) show the results for the velocity separation, which is the dominant component of physical separation between the galaxies in our close-pairs compared to the projected separation on the sky. Therefore, we use it as a proxy for closeness of our galaxy pairs. Interestingly, we see an excess of overmassive SMBHs (with up to a factor of 10 difference between M$_\text{SE}$ and M$_\text{SE,b}$) for closely separated pairs with $\Delta v \lesssim 100$ km s$^{-1}$. We also see a slight negative correlation with the M$_\text{SE}$/M$_\text{SE,b}$ and $\Delta v$ with a Pearson's-r of -0.23. We do not expect velocity separation to correlate with M$_\text{b}$, and thus the partial correlation coefficient only decreases by 0.07. Therefore, we can conclude that the SMBH mass offset depends on the pair separation, with overmassive SMBHs found in more closely separated pairs in velocity space. This trend could possibly be explained by physical processes in galaxy mergers, which we discuss in Section \ref{agndiscussion}.

Ultimately, when applying bulge mass scaling relations to AGN in the secondary bulge, the SMBH mass can be over or underestimated by orders of magnitude depending on the bulge mass ratio and close-pair separation. This result is only based on SMBHs that are actively accreting mass, which could play a role in the offsets we see for these galaxies. In order to understand the implications of scaling relations on galaxy mergers more broadly, we compare different scaling relations with the full sample of galaxy close-pairs.  

\subsection{Difference between M$_\bullet-$M$_\text{b}$ and M$_\bullet-\sigma$ for galaxies in close-pairs} \label{closepairs_results}

\subsubsection{SMBH mass ratio $q_\bullet$} \label{section:q}

\begin{figure*}[ht]
    \centering
    \includegraphics[width=0.9\textwidth]{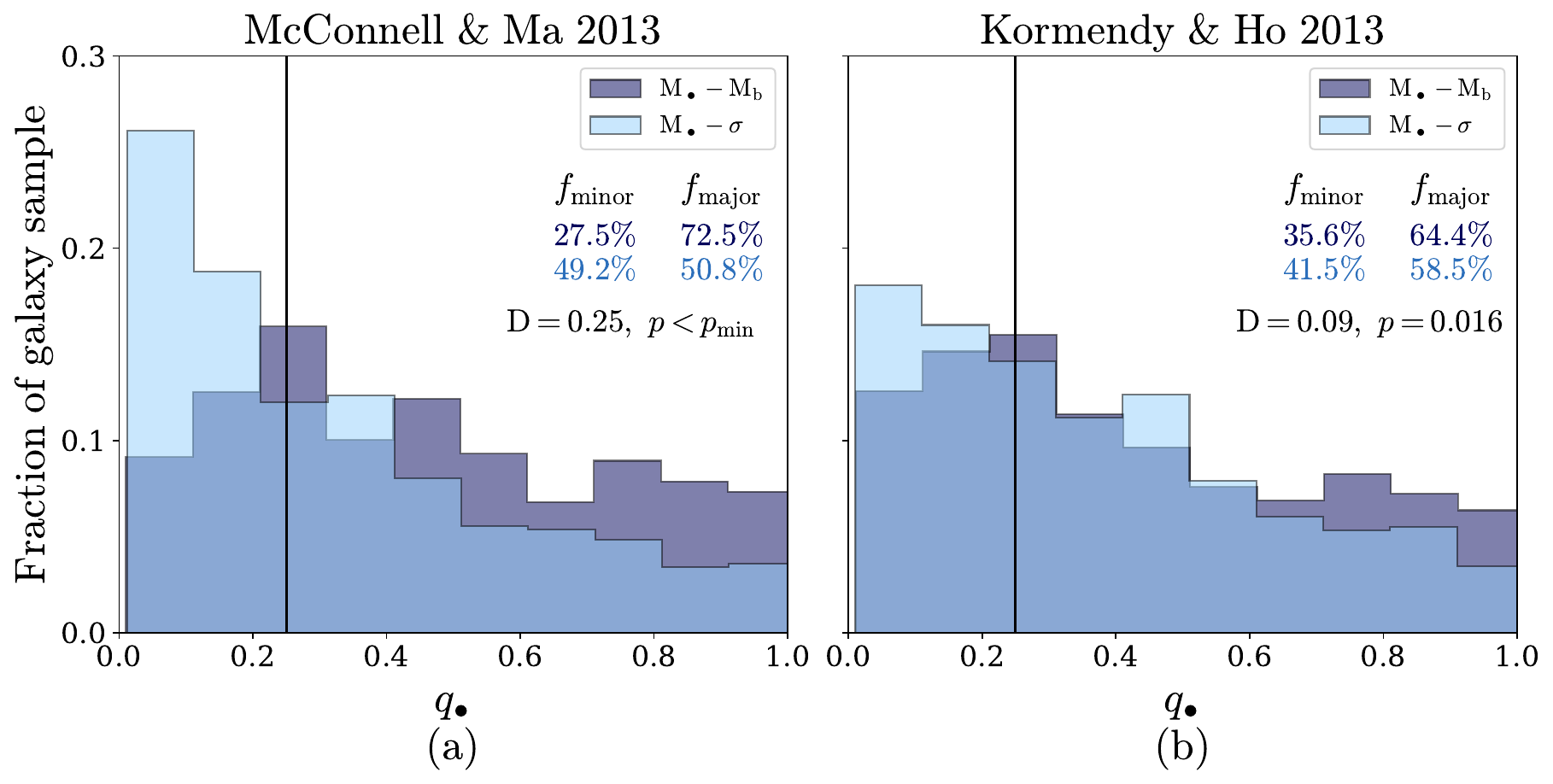}
    \caption{Normalized histograms showing the distribution of SMBH mass ratio $q_\bullet$ (Equation \ref{q}) calculated from M$_\bullet-\sigma$ (light blue) and M$_\bullet-$M$_\text{b}$ (dark blue). Figure (a) uses M$_\bullet-$M$_\text{b}$ and M$_\bullet-\sigma$ from \cite{mm13}, and Figure (b) uses M$_\bullet-$M$_\text{b}$ and M$_\bullet-\sigma$ from \cite{kh13}. The solid black line at $q_\bullet = 0.25$ divides major and minor SMBH mergers, and the relative fraction of major ($f_{\mathrm{major}}$) and minor ($f_{\mathrm{minor}}$) SMBH mergers predicted by each scaling relation are shown as percentage of the total close-pair sample. The K-S statistic (D) between the distributions and the associated p-values are also shown. D is greater in Figure (a), indicating a larger statistical deviation in the $q_\bullet$ distributions from the \cite{mm13} scaling relations (likely a result of the differences in morphology and M$_\bullet$ dynamic range described in Section \ref{pairs}). The M$_\bullet-$M$_\text{b}$ scaling relations predict a greater fraction of major SMBH pairs than M$_\bullet-\sigma$.\label{fig:qbh_distributions}}
\end{figure*}
\begin{figure*}[ht]
    \centering
    \includegraphics[width=0.9\textwidth]{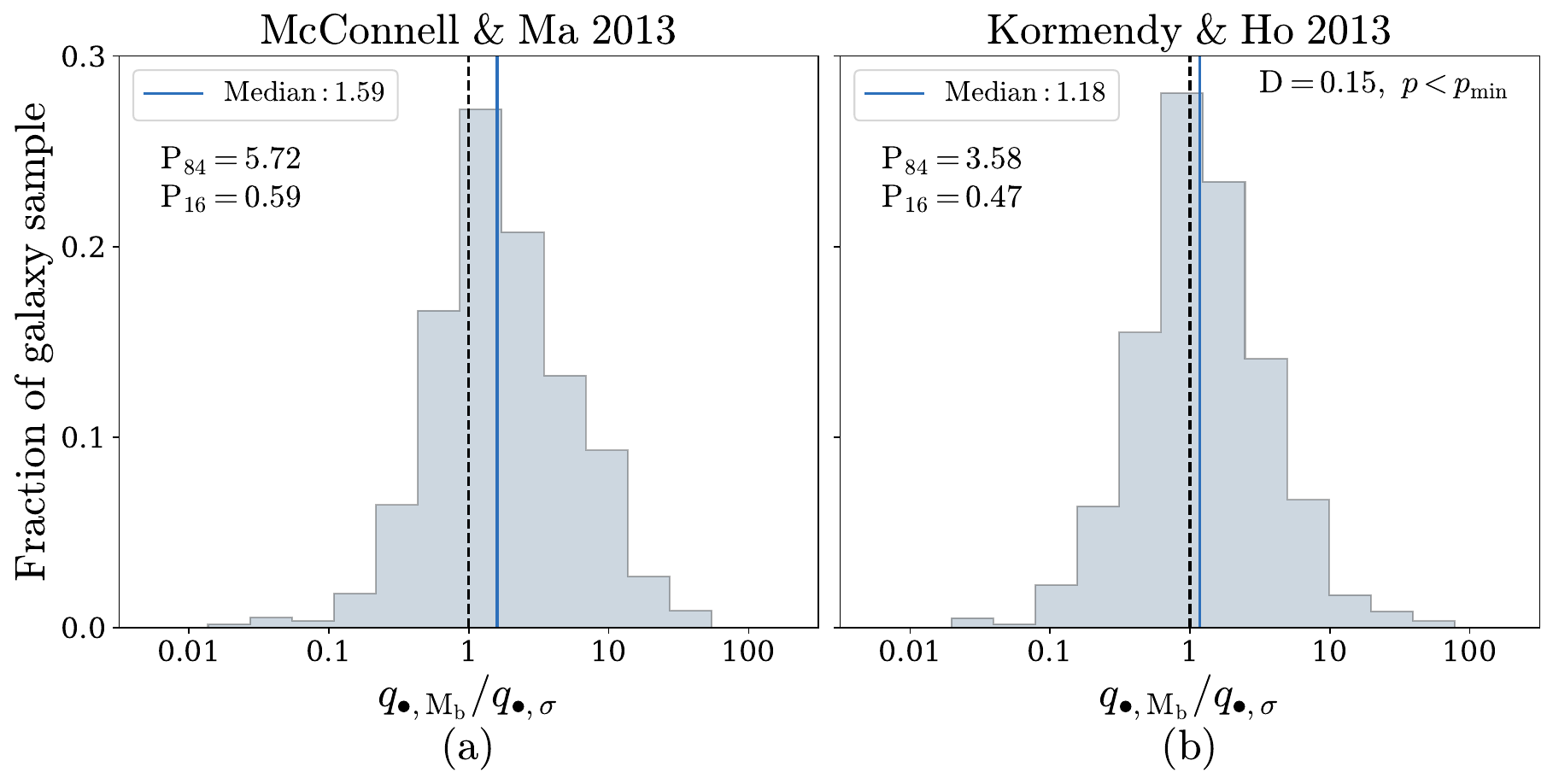}
    \caption{Normalized histograms showing the offset between $q_\bullet$ calculated by M$_\bullet-$M$_\text{b}$ and M$_\bullet-\sigma$ ($q_{\bullet,\text{M}_{b}}/q_{\bullet,\sigma}$) for individual close-pairs of galaxies. Figure (a) uses M$_\bullet-$M$_\text{b}$ and M$_\bullet-\sigma$ from \cite{mm13}, and Figure (b) uses M$_\bullet-$M$_\text{b}$ and M$_\bullet-\sigma$ from \cite{kh13}. The dashed line at 1 indicates perfect agreement, and the solid line is the median of the offset distribution. The $84^{\mathrm{th}}$ and  $16^{\mathrm{th}}$ percentiles of the distributions are also given. The K-S statistic and p-value between the distributions in Figures (a) and (b) are shown in the upper right hand corner of Figure (b). The M$_\bullet-$M$_\text{b}$ scaling relations predict mass ratios closer to 1:1 compared to M$_\bullet-\sigma$. \label{fig:qbh_offsets}}
\end{figure*}

We can use black hole-host-galaxy scaling relations on our sample of close-pairs to calculate SMBH mass ratio $q_\bullet$. In practice, this is a key way to infer what the galaxy merger population implies about the population of merging SMBHs. 

Scaling relations are the most widely applicable method of estimating SMBH mass, so they are well-suited to derive observational constraints on the SMBH binary merger population. However, the implications of using one scaling relation over another to calculate $q_\bullet$ have so far been unexplored. In this paper, we calculate $q_\bullet$ using M$_\bullet-$M$_\text{b}$ and M$_\bullet-\sigma$ from \cite{mm13} and \cite{kh13} for our sample of close galaxy pairs outlined in Section \ref{pairs} (for this section, we strictly use measured M$_\text{b}$ and $\sigma$ instead of inferred). We then compare between each set of M$_\bullet-$M$_\text{b}$ and M$_\bullet-\sigma$ scaling relations, analyzing the overall fraction of major vs. minor SMBH mergers implied by $q_\bullet$ in Figure \ref{fig:qbh_distributions} and the individual offsets between different $q_\bullet$ calculations in Figure \ref{fig:qbh_offsets}.

Figure \ref{fig:qbh_distributions} shows the distribution of SMBH mass ratios calculated by the scaling relations from Table \ref{tab:scalingrel}. Figure \ref{fig:qbh_distributions}(a) shows there is a significant difference in the relative fraction of major and minor SMBH pairs predicted by the \cite{mm13} scaling relations, where M$_\bullet-\sigma$ predicts that half ($\sim49\%$) of close galaxy pairs have minor SMBH pairs, and M$_\bullet-$M$_\text{b}$ predicts a majority ($\sim73\%$) have major SMBH pairs. There is still a difference in major and minor SMBH pair fractions between the \cite{kh13} scaling relations in Figure \ref{fig:qbh_distributions}(b), though both relations predict a majority of major SMBH pairs, with M$_\bullet-\sigma$ predicting $\sim60\%$ of the galaxy sample have major pairs and M$_\bullet-$M$_\text{b}$ predicting $\sim65\%$ have major pairs. 

We perform a two-sample Kolmogorov-Smirnov (K-S) test for the statistical divergence in the $q_\bullet$ distributions from M$_\bullet-$M$_\text{b}$ and M$_\bullet-\sigma$ (Figure \ref{fig:qbh_distributions}). Our test statistic is D $ =0.25$ between the \cite{mm13} $q_{\bullet,\sigma}$ and $q_{\bullet,\text{M}_\text{b}}$ distributions and D $ = 0.09$ for the \cite{kh13} distributions, confirming that the divergence between $q_{\bullet,\sigma}$ and $q_{\bullet,\text{M}_\text{b}}$ is more significant between the \cite{mm13} relations, with \cite{kh13} giving $q_\bullet$ values from different scaling relations that are generally consistent with each other.

This closer agreement between the \cite{kh13} scaling relations is expected, as they only include galaxies with classical bulges. The \cite{mm13} scaling relations, however, include pseudo-bulges in M$_\bullet-\sigma$ but not M$_\bullet-$M$_\text{b}$. This results in the \cite{mm13} M$_\bullet-\sigma$ covering a larger dynamic range of M$_\bullet$, allowing for lower mass SMBHs and smaller binary mass ratios. These differences, as shown in Figure \ref{fig:qbh_distributions}, propagate through the $q_\bullet$ calculations and result in different local ($z \leq 0.2$) populations of SMBH pairs, where the greater dynamical range in \cite{mm13} M$_\bullet-\sigma$ allows for lower mass SMBHs at a given $\sigma$ and more extreme binary mass ratios overall. 

While Figure \ref{fig:qbh_distributions} shows how scaling relations change the overall population of galaxy mergers hosting minor vs. major SMBH pairs, Figure \ref{fig:qbh_offsets} shows how scaling relations return different $q_\bullet$ values for the same galaxy pair. As in Figure \ref{fig:qbh_distributions}, Figure \ref{fig:qbh_offsets} shows a smaller difference in $q_\bullet$ between the \cite{kh13} scaling relations than \cite{mm13}. Both sets of scaling relations show an offset toward higher values of $q_\bullet$ from M$_\bullet-$M$_\text{b}$ than M$_\bullet-\sigma$, with larger tails toward offsets $q_{\bullet,\text{M}_\text{b}}/q_{\bullet,\sigma} > 1$ that extend to 1-2 orders of magnitude (accounting for $3-7\%$ of pairs). The median $q_{\bullet,\text{M}_\text{b}}/q_{\bullet,\sigma}$ is 1.6 for \cite{mm13} and 1.2 for \cite{kh13}. This implies that for a given galaxy merger, M$_\bullet-$M$_\text{b}$ assumes that equal-mass bulges imply equal-mass black holes, but M$_\bullet-\sigma$ could predict a more minor/unequal-mass SMBH pair even in galaxies with comparable stellar bulges. The opposite happens as well, where M$_\bullet-$M$_\text{b}$ predicts more unequal SMBH masses than M$_\bullet-\sigma$ based on unequal bulge masses. This offset occurs in fewer galaxy pairs and extends to $\sim$1 order of magnitude for our sample, though $\sim1\%$ of pairs have a $q_{\bullet,\text{M}_\text{b}}$ less than $q_{\bullet,\sigma}$ by over an order of magnitude.

Overall, our results show significantly diverging $q_\bullet$ values between the M$_\bullet-$M$_\text{b}$ and M$_\bullet-\sigma$ relations. This divergence has implications on the evolution of the SMBH pair throughout the merger, even though our pair sample is only able to resolve those at kpc-scale separations. There is evidence from both dual-AGN modeling \citep[e.g.,][]{yu2011,capelo2015,capelo2017} and observations \citep[e.g.,][]{shen2011,comerford2015,barrows2023} to support stronger and more rapid SMBH mass growth in the secondary galaxy, especially after the first pericenter passage. Which SMBH-host-galaxy scaling relation is used to derive SMBH masses therefore directly affects the resulting evolution of the SMBH pair in a galaxy merger.

\subsubsection{Chirp mass $\mathcal{M}$}

\begin{figure*}[ht]
    \centering
    \includegraphics[width=0.9\textwidth]{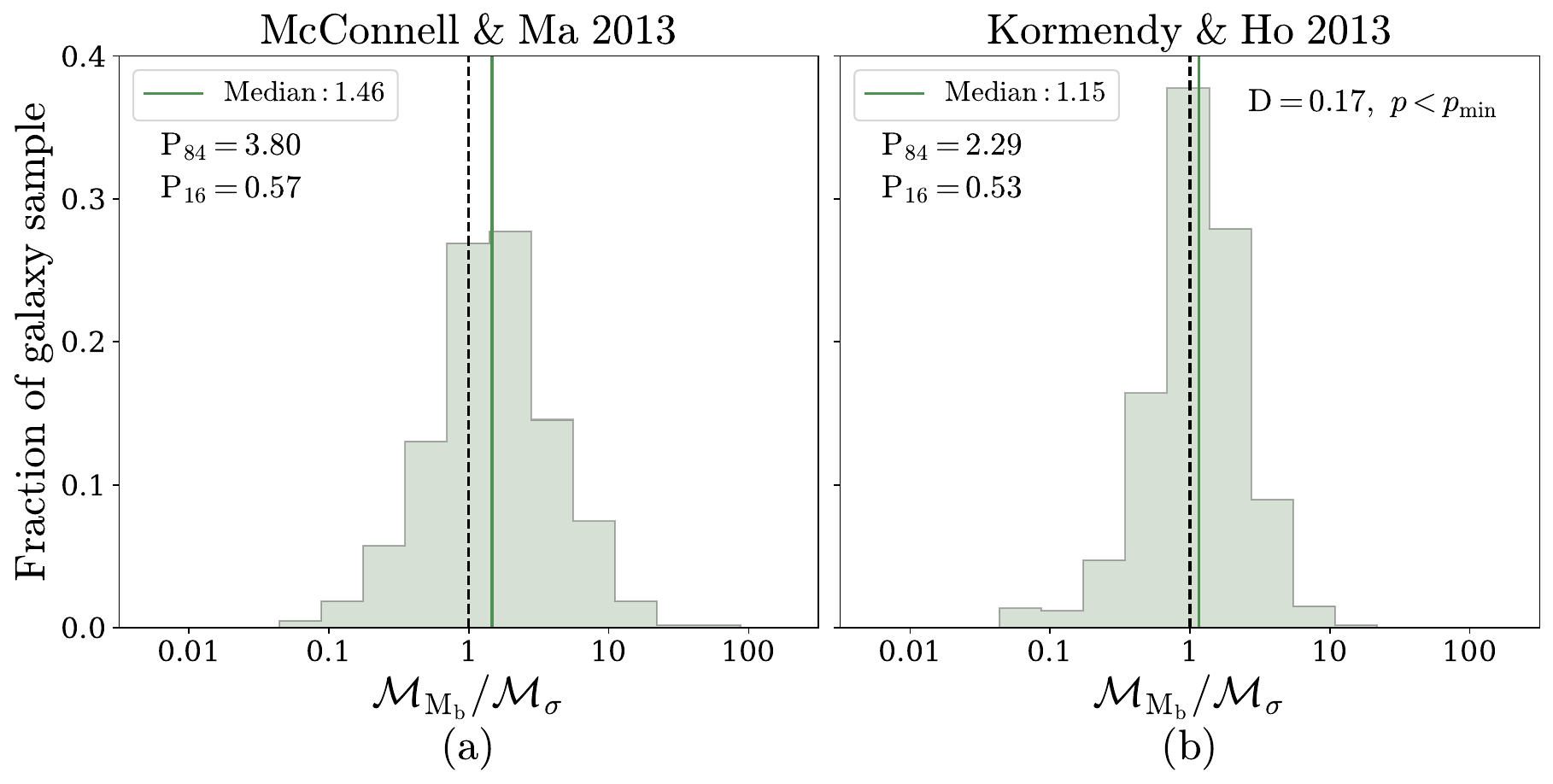}
    \caption{Normalized histograms showing the offset between chirp mass $\mathcal{M}$ (Equation \ref{chirpmass}) calculated from M$_\bullet-$M$_\text{b}$ and from M$_\bullet-\sigma$ ($\mathcal{M}_{\text{M}_\text{b}}/\mathcal{M}_{\sigma}$) for individual close galaxy pairs. Figure (a) uses M$_\bullet-$M$_\text{b}$ and M$_\bullet-\sigma$ from \cite{mm13} and Figure (b) uses M$_\bullet-$M$_\text{b}$ and M$_\bullet-\sigma$ from \cite{kh13}. The dashed line at 1 indicates perfect agreement, and the solid line is the median of the offset distribution. The $84^{\mathrm{th}}$ and  $16^{\mathrm{th}}$ percentiles of the distributions are also given, with the K-S test statistic and p-value between the distributions in Figure (a) and (b) in the upper right hand corner of Figure (b). The M$_\bullet-$M$_\text{b}$ scaling relations predict higher chirp masses compared to M$_\bullet-\sigma$.\label{fig:chirp_offsets}}
\end{figure*}

So far, we find evidence to suggest that M$_\bullet-$M$_\text{b}$ and M$_\bullet-\sigma$ diverge in their predictions for both the SMBH mass ratio $q_\bullet$ (as shown in Section \ref{section:q}) and individual SMBH masses. The chirp mass $\mathcal{M}$, which factors into the gravitational-wave amplitude produced by SMBH mergers, depends on both the mass ratio and total mass of the binary (Equation \ref{chirpmass}). Greater values of $\mathcal{M}$ lead to greater gravitational-wave amplitudes. We use our sample to calculate the offset in $\mathcal{M}$ between M$_\bullet-$M$_\text{b}$ and M$_\bullet-\sigma$ and demonstrate how the individual divergence in $q_\bullet$ and M$_\bullet$ combine and propagate through the $\mathcal{M}$ calculation. 

Figure \ref{fig:chirp_offsets} shows the ratio between $\mathcal{M}$ calculated from M$_\bullet-$M$_\text{b}$ and M$_\bullet-\sigma$. Figure \ref{fig:chirp_offsets}(a) shows the offset between the \cite{mm13} scaling relations, where the distribution is skewed away from the one-to-one line and toward higher offsets. This means that for the \cite{mm13} scaling relations, M$_\bullet-$M$_\text{b}$ is predicting larger chirp masses than M$_\bullet-\sigma$ by of a factor $1.5$ on average, with $2\%$ of galaxies offset by greater than 1 order of magnitude. Figure \ref{fig:chirp_offsets}(b) shows the offset between the \cite{kh13} scaling relations, which agree on chirp mass far more than the \cite{mm13} scaling relations with a slight offset toward greater chirp masses from M$_\bullet-$M$_\text{b}$ by an average factor of 1.2, with only $0.1\%$ of galaxies with $\mathcal{M}_{\text{M}_\text{b}}/\mathcal{M}_{\sigma}$ greater than 1 order of magnitude.

The chirp mass results show that M$_\bullet-$M$_\text{b}$ tends to predict larger chirp masses than M$_\bullet-\sigma$ for most galaxy pairs in our sample. The skew is larger for the \cite{mm13} scaling relations, in accordance with the $q_\bullet$ results (Figures \ref{fig:qbh_distributions}, \ref{fig:qbh_offsets}). The offset in chirp mass adds to existing evidence that differences encoded in scaling relations significantly impact our interpretation of gravitational-wave signals from SMBH mergers \citep[e.g.,][]{simon2016,simon2023,cayenne2023,matt2025}.

\subsection{Using M$_{\mathrm{b,inf}}$ and $\sigma_\mathrm{inf}$ to calculate $q_\bullet$ and $\mathcal{M}$}\label{infresults}
\begin{figure*}[ht]
    \centering
    \includegraphics[width=0.9\textwidth]{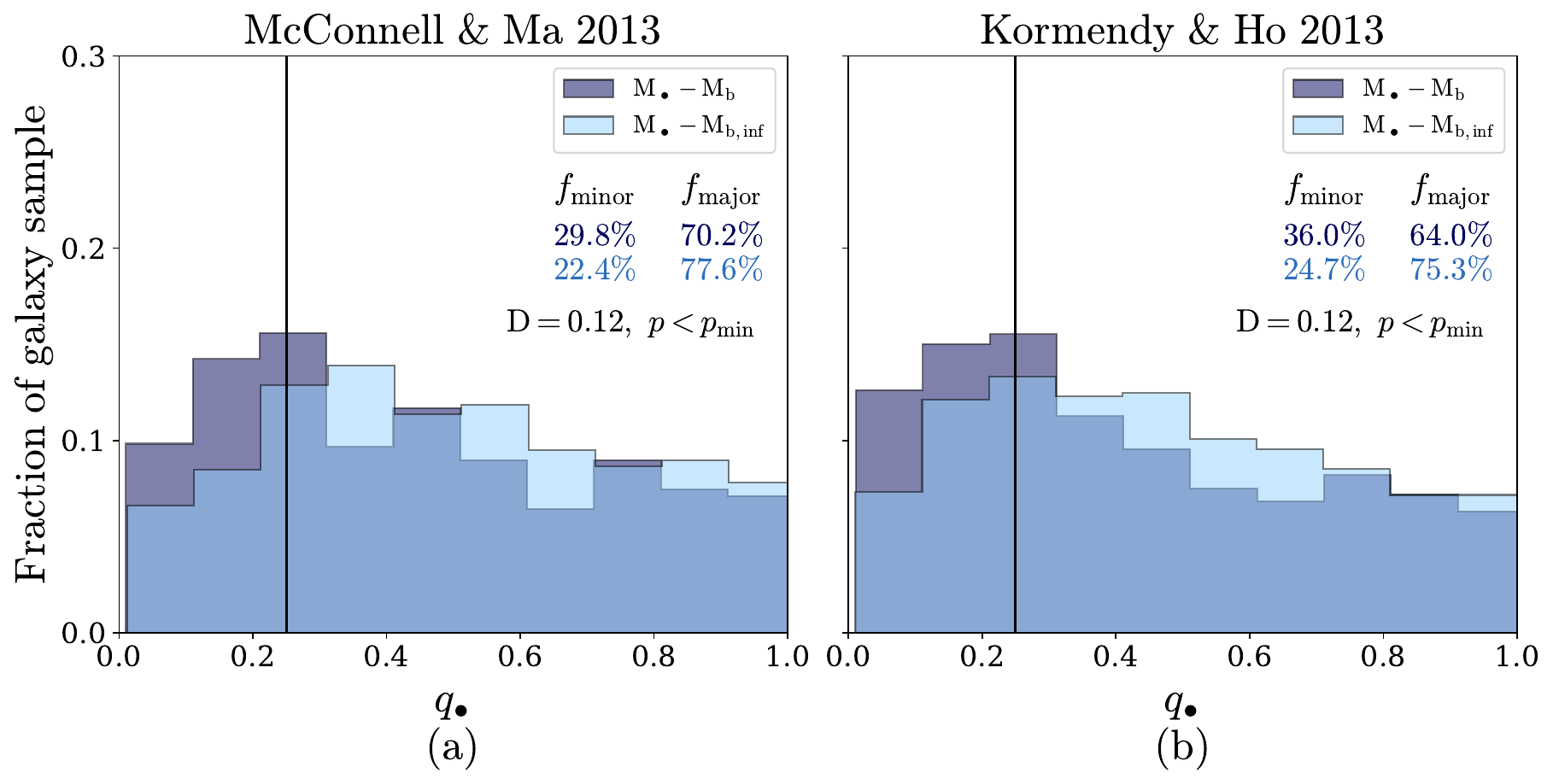}
    \includegraphics[width=0.9\textwidth]{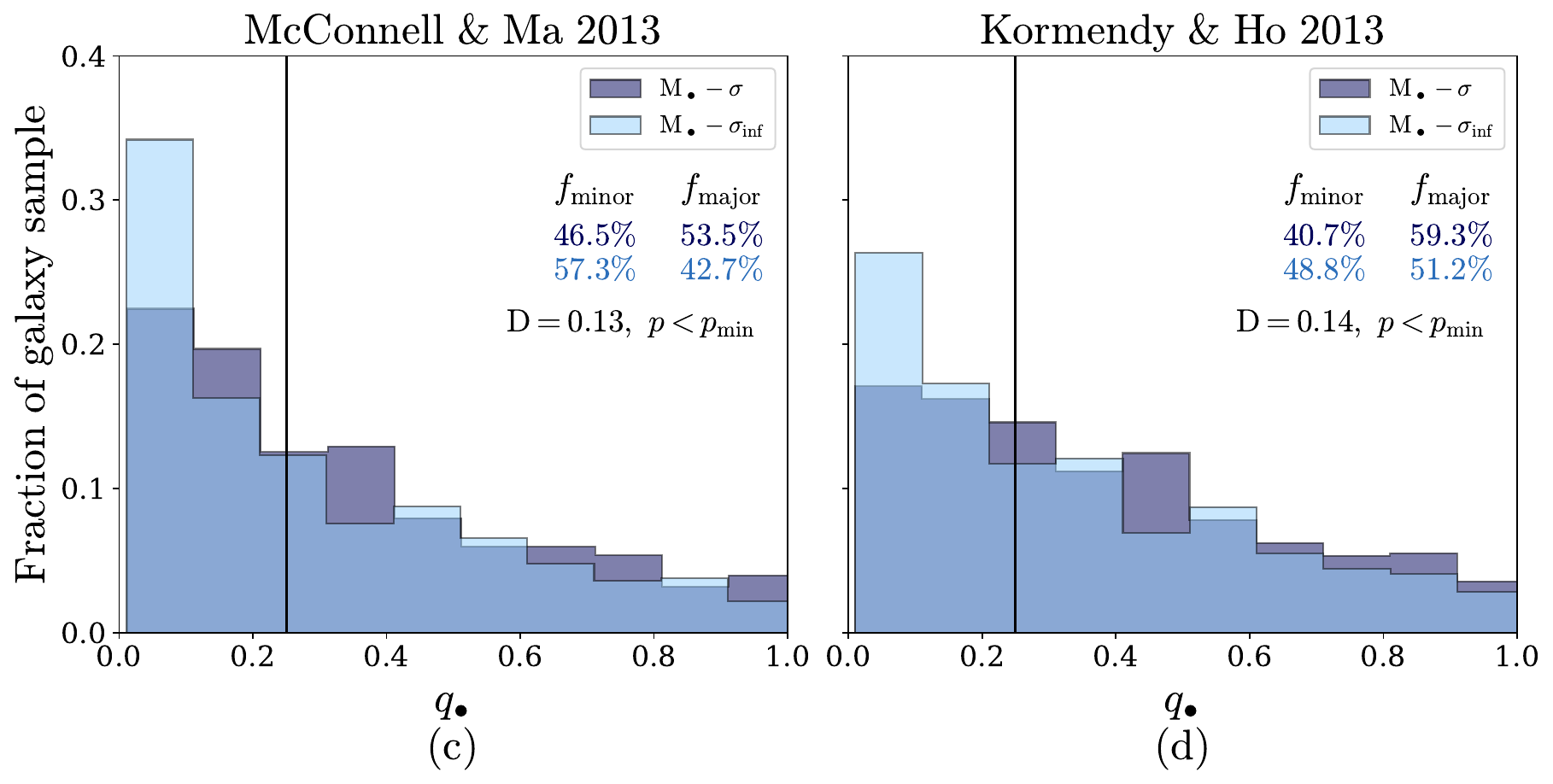}
    \caption{Normalized histograms showing the distribution of SMBH mass ratio $q_\bullet$ (Equation \ref{q}) calculated from inferred quantities M$_\bullet-$M$_{\mathrm{b,inf}}$ (Equations \ref{fb1}, \ref{fb2}) and M$_\bullet-\sigma_\text{inf}$ (Equation \ref{inferred veldisp}) in light blue and M$_\bullet-$M$_\text{b}$ and M$_\bullet-\sigma$ in dark blue. Figures (a) and (c) use the \cite{mm13} relation and Figures (b) and (d) use \cite{kh13}. The solid black line at $q_\bullet = 0.25$ divides major and minor SMBH mergers, and the relative fraction of major ($f_{\mathrm{major}}$) and minor ($f_{\mathrm{minor}}$) SMBH mergers predicted by each scaling relation are shown as percentage of the total close-pair sample. The K-S test statistic and p-values between the light and dark blue distributions are in the upper right corners of each figure. M$_\bullet-$M$_{\mathrm{b,inf}}$ predicts a greater fraction of major SMBH mergers than M$_\bullet-$M$_\text{b}$, while M$_\bullet-\sigma_\text{inf}$ predicts a greater fraction of minor mergers than M$_\bullet-\sigma$. \label{fig:qbh_inf_distributions}}
\end{figure*}

To determine how inferring M$_\text{b}$ will impact our results, we reproduce the analysis from Section \ref{closepairs_results} for M$_\bullet-$M$_\text{b}$ and M$_\bullet-$M$_{\mathrm{b,inf}}$. We use the updated M$_{\mathrm{b,inf}}$ prescription from Section \ref{infbulge} as it agrees the best with M$_\text{b}$ from \cite{mendel13}. Figures \ref{fig:qbh_inf_distributions} (a) and (b) show that inferring M$_\text{b}$ does seem to bias the overall distribution of $q_\bullet$, with inferred bulge mass predicting $\sim 7 \%$ more major SMBH mergers overall with \cite{mm13} (Figure \ref{fig:qbh_inf_distributions}(a)), and $11\%$ with \cite{kh13} (Figure \ref{fig:qbh_inf_distributions}(b)). Although this is a modest increase in the fraction of major mergers, the bias will artificially push the M$_\bullet-$M$_\text{b}$ mass ratios even higher with respect to the mass ratios from M$_\bullet-\sigma$. 

As for the impact of inferring $\sigma$ from the virial theorem, Figures \ref{fig:qbh_inf_distributions} (c) and (d) show that the overall $q_\bullet$ distribution will shift toward a greater fraction of minor mergers. For the \cite{mm13} M$_\bullet-\sigma$, using $\sigma_\mathrm{inf}$ increases the overall fraction of minor mergers by $11\%$ (Figure \ref{fig:qbh_inf_distributions}(c)). Using $\sigma_\mathrm{inf}$ with the \cite{kh13} M$_\bullet-\sigma$ increases the minor merger fraction by $8\%$ (Figure \ref{fig:qbh_inf_distributions}(d)).

Overall, our results show that choosing to infer both $\sigma$ and M$_\text{b}$ will widen the divergence in the estimated $q_\bullet$ distributions from M$_\bullet-$M$_\text{b}$ and M$_\bullet-\sigma$. This will cause significantly different SMBH binary population inferences in terms of the relative fraction of major and minor SMBH mergers, and further affect any parameters (such as chirp mass) that rely on $q_\bullet$ estimates.

\section{Discussion} \label{discussion}
\subsection{SMBH mass growth in close galaxy pairs hosting AGN} \label{agndiscussion}
In this paper, we use the scaling relation-independent single epoch virial SMBH mass to observationally test SMBH mass growth relative to the stellar bulge in galaxy close-pairs. Our results from Section \ref{MSE results} show evidence of elevated SMBH mass growth relative to the bulge for AGN that reside in the galaxy that hosts the secondary bulge and in more closely separated pairs. We explore findings from previous studies, pertaining to both observations and simulations of galaxy mergers, that provide possible astrophysical explanations for our results here. 

It is known that galaxy mergers can induce AGN activity via gravitational torques that funnel gas to the centers of galaxies \citep[e.g.,][]{shlosman1990,mihos1996,barnes1996}, with elevated activity observed in close-pairs out to projected separations of $40 - 60$ kpc \citep{ellison11,fu2018,steffen2023} and with AGN excess peaking at SMBH separations $<10$ kpc \citep{stemo2021}. This agrees with our results in Figure \ref{fig:other_correlations}(b), where the most overmassive SMBHs exist at the closest separations between pairs. 

There is evidence that the AGN in secondary galaxies in both major and minor mergers accrete with higher Eddington ratios than the primary galaxy for pairs with dual AGN, as found in hydrodynamical simulations \citep[e.g.,][]{vanwassenhove2012,capelo2015,steinborn2016,capelo2017,volonteri2022} and dual AGN observations \citep[e.g.,][]{comerford2015,barrows2023}. The presence of overmassive SMBHs in secondary bulges in our sample agrees with these findings. As for major galaxy mergers, there is a greater excess of AGN overall \citep[e.g.,][]{ellison11,comerford2024}, where simulations show both galaxies are strongly perturbed by the merger and experience significant SMBH mass growth \citep[e.g.,][]{capelo2015,capelo2017,volonteri2022}. Our results find that major galaxy mergers preferentially host overmassive SMBHs relative to their bulges. 

The same gas inflows that can activate the central black holes of merging galaxies can also be used to form stars and grow stellar bulges. Star formation (SF) enhancement is accordingly observed in mergers out to separations $<150$ kpc \citep{patton13} with the SF enhancement most prominent at separations $<30$ kpc. Major mergers experience stronger SF bursts overall, in both the primary and secondary galaxies \citep[e.g.,][]{davies2015,steffen2021}. However, observations also show that secondary galaxies in minor mergers experience SF suppression as the primary galaxy strips away their gas, and this suppression is most prominent at separations $<30$ kpc. Violent relaxation of disk stars and nuclear starbursts can both trigger bulge mass growth in mergers. Bulge assembly through major mergers appears to dominate other mass growth channels in the most massive bulges, with disk instabilities and merger-induced star formation contributing more for lower mass bulges \citep[e.g.,][]{hopkins2010,bell2017,husko2022}. 

Our sample of AGN in galaxy close-pairs includes projected separations out to $100$ kpc, with $25\%$ of the sample $<40$ kpc, where both merger-induced AGN triggering and SF are most relevant. With that in mind, our results may arise due to the differential merger-induced growth between the stellar mass and the SMBH. Figure \ref{fig:other_correlations} and Table \ref{tab:offsets} show that most SMBHs in secondary stellar bulges are overmassive relative to M$_\bullet-$M$_\text{b}$. This supports existing evidence from dynamical SMBH masses that SMBH growth outpaces the bulge during a merger \citep{medling2015}. The gap between the true SMBH mass and the host-galaxy prediction may come down to a difference in timescales, where merger-induced bulge mass growth in the secondary galaxy is not fast enough to keep scaling relations intact during ongoing mergers as the central SMBH grows through accretion. Cosmological simulations tend to show less massive SMBHs relative to their bulges at high redshift ($z\sim2$) \citep[e.g.,][]{steinborn2016,puerto-sanchez2024} for both primary and secondary AGN hosts, though these studies compare to local scaling relations to which these simulations are already calibrated.

This study observationally shows for the first time that the gap between SMBH mass and M$_\bullet-$M$_\text{b}$ predictions depends on the stellar mass ratio of the galaxy pair. Evidence of suppressed star formation combined with faster SMBH growth in the secondary galaxies of minor mergers, which we have discussed here, could explain the mismatch between SMBH predictions for these galaxies. The SF suppression and SMBH enhancement both scale with close-pair separation, which also could explain why we find SMBHs becoming more overmassive as the galaxy pairs become closer in 3-D separation. 

\subsection{The implications of using scaling relations to calculate $q_\bullet$ and $\mathcal{M}$}
In addition to exploring the relative growth of the SMBH and the bulge for AGN hosts, we present results for the full sample of galaxy close-pairs in SDSS and compare $q_\bullet$ and $\mathcal{M}$ when calculated using M$_\bullet-$M$_\text{b}$ and M$_\bullet-\sigma$. We find that overall, M$_\bullet-$M$_\text{b}$ predicts higher values of both $q_\bullet$ and $\mathcal{M}$ for most of our close-pairs, with the greatest offset between the \cite{mm13} scaling relations. This reflects the fact that the \cite{mm13} M$_\bullet-\sigma$ relation is fit for the largest dynamic range in M$_\bullet$ by including low-mass and late-type galaxies, and thus predicts a wider range of M$_\bullet$ estimates. This allows for more unequal mass ratios and thus smaller $q_\bullet$ values. There is also the possibility that M$_\bullet-\sigma$ in general reflects the more fundamental relation between SMBHs and their host galaxies, and the M$_\bullet-$M$_\text{b}$ assumption follows from the projection of this relation or from non-causal effects such as hierarchical merging \citep[e.g.,][]{peng2007,jahnke2011,vandenbosch2016,dn19}. 

The results shown in Figures \ref{fig:qbh_distributions} - \ref{fig:chirp_offsets} are particularly relevant to astrophysical gravitational-wave predictions. The difference in $\mathcal{M}$ with scaling relations will directly change the inferred amplitude of the stochastic GWB from PTAs and individual GW signals from LISA, with higher chirp masses corresponding to higher amplitudes. In addition to the dependence of GW signals on $\mathcal{M}$, differences in $q_\bullet$ calculations also affect other several parameters relevant to the GWB. It has been shown that the evolution of $q_\bullet$ throughout the merger strongly depends on the initial $q_\bullet$ \citep{capelo2015}, which we effectively study here as we consider pairs with kpc-scale separations. The initial $q_\bullet$ also influences the coalescence timescale for SMBH binaries \citep{khan2012a,berczik2022,holley-bockelmann2025}, where unequal-mass binaries are faster to coalesce and the number of detectable sources can increase with a greater population of minor SMBH mergers \citep{siwek2024}. The initial mass ratio can also influence SMBH binary hardening and the evolution of orbital parameters such as semi-major axis and eccentricity \citep{siwek2020,siwek2023}, where binaries with  $q_\bullet \gtrsim 0.3$ are more likely to evolve toward coalescence and binaries with lower mass ratios are more likely to stall in their orbit due to torques from the circumbinary disk. 

We show here that the adoption of a particular scaling relation to construct a mass function changes the initial $q_\bullet$ distribution, which will change the overall gravitational-wave signal from SMBHBs. Given the wealth of evidence from the literature to support M$_\bullet-\sigma$ as the fundamental scaling relation \citep[e.g.,][]{peng2007,jahnke2011,hirschmann2010,wake2012,king2015,vandenbosch2016,dn19,marsden2020,shankar2025,Huber2025}, this could mean that the relative abundance of minor mergers is significantly underestimated by studies that use M$_\bullet-$M$_\text{b}$.

\section{Conclusion} \label{conclusion}
In this study, we have used the broad-line AGN, bulge masses, velocity dispersions, and close galaxy pairs at $0.02 \leq z \leq 0.2$ from SDSS DR7 as a laboratory to test black hole-host-galaxy scaling relations in mergers. With our well-studied and mass-complete sample of galaxy mergers and updated single-epoch mass prescriptions that are scaling relation-independent, we were able to investigate how scaling relations diverge in their predictions for the black hole mass ratio and chirp mass for low-redshift close-pairs. Our findings show that:

\begin{itemize}
    \item We fit a new power-law scaling relation between single-epoch virial SMBH mass and bulge stellar mass for non-merging galaxies that host broad line AGN as our fiducial nonmerger based M$_\bullet-$M$_\text{b}$ (Figure \ref{fig:mse_mbulge}). We calculate the ratio between the ground truth single-epoch SMBH mass for AGN hosts in galaxy pairs and the SMBH mass predicted by M$_{\mathrm{SE}}-$M$_\text{b}$ and explore its dependence on the bulge stellar mass ratio $q_\text{b}$ and the pair line-of-sight velocity separation $\Delta v$ (Figure \ref{fig:other_correlations} and Table \ref{tab:offsets}). We find that AGN residing in secondary bulges of with small mass ratios ($0.01 < q_\text{b} \leq 0.1$) have SMBHs that are overmassive compared to M$_{\mathrm{SE}}-$M$_\text{b}$ by a factor of two on average. We also find overmassive SMBHs by a factor of two in closely separated galaxy pairs with $\Delta v < 100$ km s$^{-1}$. 
    
    \item We expanded our study to include all galaxy pairs regardless of whether they host an AGN. Without a universal ground truth estimate of SMBH mass, we compare the \cite{mm13} and \cite{kh13} M$_\bullet-$M$_\text{b}$ and M$_\bullet-\sigma$ scaling relations with a sample of $\sim600$ close-pairs with a merger probability greater than 50$\%$. We examine the difference between their predictions of the SMBH mass ratio $q_\bullet$ distribution (Figure \ref{fig:qbh_distributions}). We find significant divergence in the $q_\bullet$ distributions between M$_\bullet-$M$_\text{b}$ and M$_\bullet-\sigma$, with the greatest difference between the \cite{mm13} scaling relations. The greater difference in \cite{mm13} likely arises due to the different SMBH mass ranges in the galaxies used to fit M$_\bullet-$M$_\text{b}$ and M$_\bullet-\sigma$, with M$_\bullet-\sigma$ extending to lower SMBH masses. For \cite{mm13}, M$_\bullet-$M$_\text{b}$ predicts $27.5\%$ of the galaxy pairs host minor (unequal-mass) SMBH pairs and $72.5\%$ major (equal-mass) pairs. \cite{mm13} M$_\bullet-\sigma$, on the other hand, predicts more equal fractions with $49\%$ minor SMBH pairs and $51\%$ major pairs- a $22\%$ increase in the relative fraction of minor mergers compared to M$_\bullet-$M$_\text{b}$. The $q_\bullet$ distributions agree better for the \cite{kh13} scaling relations, but still show an offset toward M$_\bullet-\sigma$ predicting $\sim6\%$ more minor SMBH pairs than M$_\bullet-$M$_\text{b}$. 

   We calculate the offset in the measured $q_\bullet$ values between M$_\bullet-$M$_\text{b}$ and M$_\bullet-\sigma$ for the same close galaxy pair (Figure \ref{fig:qbh_offsets}). We find that M$_\bullet-$M$_\text{b}$ measures a larger $q_\bullet$ than M$_\bullet-\sigma$ by an average factor of 1.6 for \cite{mm13} and 1.2 for \cite{kh13}.
    
    \item To further understand the impact of the choice of scaling relation on gravitational-wave predictions, we investigated how the differences between scaling relations affect the calculation of the chirp mass for SMBHs in galaxy close-pairs. The offset in $\mathcal{M}$ is more significant between the \cite{mm13} scaling relations than the \cite{kh13} scaling relations, though both show an offset toward M$_\bullet-$M$_\text{b}$ implying higher $\mathcal{M}$ than M$_\bullet-\sigma$. \cite{mm13} shows an offset by a factor of 1.5 and \cite{kh13} shows an offset by a factor of 1.2.

    \item Lastly, we take commonly used prescriptions to infer M$_\text{b}$ and $\sigma$ and use these values to calculate $q_\bullet$ for our galaxy close-pair sample. We find that inferring M$_\text{b}$ and $\sigma$ widens the gap between the relative major/minor SMBH pair fractions predicted by M$_\bullet-$M$_\text{b}$ and M$_\bullet-\sigma$, with M$_{\mathrm{b,inf}}$ predicting $5-7\%$ more major mergers than bulge-disk decomposed M$_\text{b}$ and $\sigma_{\mathrm{inf}}$ predicting $8-11\%$ more minor mergers than spectroscopic $\sigma$.
\end{itemize}

Observational constraints on SMBH binary populations are necessary to make realistic predictions of their gravitational-wave signatures, but come with their own assumptions that need to be investigated independently. Our results show that the differences in local scaling relations significantly affect observational constraints on the SMBH binary population, and local scaling relations may skew toward underestimating the true SMBH mass in secondary stellar bulges. It is therefore important to take these biases into account whenever scaling relations are used to construct a model population of SMBH binaries. This is the case when interpreting stochastic gravitational-wave signals from PTAs or LISA, calibrating cosmological simulations, or comparing high-redshift JWST black holes to local scaling relations.

In the future, we plan to move toward a comprehensive method of modeling SMBH binaries from galaxy observations that go beyond the assumption that the stellar components of galaxies reflect the nature of the SMBHs residing in their centers. As we have shown, this assumption can severely skew the demographics we infer for SMBH pairs in the universe. 

\section*{Acknowledgments}
M.C.H. and J.M.C. acknowledge support from NSF AST-1847938 and NSF AST-2510894.

Funding for the Sloan Digital Sky Survey (SDSS) has been provided by the Alfred P. Sloan Foundation, the Participating Institutions, the National Aeronautics and Space Administration, the National Science Foundation, the U.S. Department of Energy, the Japanese Monbukagakusho, and the Max Planck Society. The SDSS Web site is http://www.sdss.org/.

The SDSS is managed by the Astrophysical Research Consortium (ARC) for the Participating Institutions. The Participating Institutions are The University of Chicago, Fermilab, the Institute for Advanced Study, the Japan Participation Group, The Johns Hopkins University, Los Alamos National Laboratory, the Max-Planck-Institute for Astronomy (MPIA), the Max-Planck-Institute for Astrophysics (MPA), New Mexico State University, University of Pittsburgh, Princeton University, the United States Naval Observatory, and the University of Washington.

This work utilized the Alpine high performance computing resource at the University of Colorado Boulder. Alpine is jointly funded by the University of Colorado Boulder, the University of Colorado Anschutz, Colorado State University, and the National Science Foundation (award 2201538).

\bibliography{references}
\bibliographystyle{aasjournal}
\end{document}